%% file: MetrologyReview_Revision.tex
\documentclass[twocolumn,prx,aps,pra,groupedaddress,superscriptaddress]{revtex4-1}
\usepackage[latin9]{inputenc}
\usepackage{color}
\usepackage{amstext}
\usepackage{amssymb}
\usepackage{amsmath}
\usepackage{amssymb}
\usepackage{txfonts}
\usepackage{textcomp}
\usepackage{graphicx}
\definecolor{myurlcolor}{rgb}{0,0,0.7}
\definecolor{myrefcolor}{rgb}{0.8,0,0}
\usepackage[unicode=true,pdfusetitle, bookmarks=true,bookmarksnumbered=false,bookmarksopen=false, breaklinks=false,pdfborder={0 0 0},backref=false,colorlinks=true, linkcolor=myrefcolor,citecolor=myurlcolor,urlcolor=myurlcolor]{hyperref}

\usepackage{dsfont}

\include{commands}



\begin{document}

\title{Precision Limits in Quantum Metrology with Open Quantum Systems}

\author{J. F. Haase}\email{jan.frhaase@gmail.com}
\affiliation{Institut f{\"u}r Theoretische Physik, Albert-Einstein-Allee 11, Universit{\"a}t Ulm, D-89069 Ulm, Germany}
\affiliation{Center for Integrated Quantum Science and Technology (IQST), Albert-Einstein-Allee 11, Universit{\"a}t Ulm, D-89069 Ulm, Germany}
\author{A. Smirne}
\affiliation{Institut f{\"u}r Theoretische Physik, Albert-Einstein-Allee 11, Universit{\"a}t Ulm, D-89069 Ulm, Germany}
\affiliation{Center for Integrated Quantum Science and Technology (IQST), Albert-Einstein-Allee 11, Universit{\"a}t Ulm, D-89069 Ulm, Germany}
\author{J. Ko{\l}ody{\'n}ski}
\affiliation{ICFO-Institut de Ci\`{e}nces Fot\`{o}niques, The Barcelona Institute of Science and Technology, 08860 Castelldefels (Barcelona), Spain}
\author{R. Demkowicz-Dobrza\'{n}ski}
\affiliation{Faculty of Physics, University of Warsaw, 02-093 Warszawa, Poland}
\author{S. F. Huelga}
\affiliation{Institut f{\"u}r Theoretische Physik, Albert-Einstein-Allee 11, Universit{\"a}t Ulm, D-89069 Ulm, Germany}
\affiliation{Center for Integrated Quantum Science and Technology (IQST), Albert-Einstein-Allee 11, Universit{\"a}t Ulm, D-89069 Ulm, Germany}

  \begin{abstract}
{The laws of quantum mechanics allow to perform measurements whose precision supersedes results predicted by classical parameter estimation theory. That is, the precision bound imposed by the central limit theorem in the estimation of a broad class of parameters, like atomic frequencies in spectroscopy or external magnetic field in magnetometry, can be overcome when using quantum probes. Environmental noise, however, generally alters the ultimate precision that can be achieved in the estimation of an unknown parameter. This tutorial reviews recent theoretical work aimed at obtaining general precision bounds in the presence of an environment. We adopt a complementary approach, where we first analyze the problem within the general framework of describing the quantum systems in terms of quantum dynamical maps and then relate this abstract formalism to a microscopic description of the system's dissipative time evolution. We will show that although some forms of noise do render quantum systems to be standard quantum limited, precision beyond classical bounds is still possible in the presence of different forms of local environmental fluctuations.}
\end{abstract}

\maketitle

\section{Introduction}
Quantum metrology is a paradigmatic example of the possible advantage provided by the use of quantum features as compared to a classical setting \cite{Giovannetti2004}. Given certain constraints, typically the number $N$ of available probes and the total experimental duration $T$, the best precision in the estimation of a given parameter is governed by the Standard Quantum Limit (SQL), a result dictated by the central limit theorem of classical statistics \cite{Kay1993}. However, exploiting the very nature of quantum states and quantum measurements allows for a different metrological bound, the so called Heisenberg limit (HL), surpassing the classically attainable precision when subject to the same constraints \cite{Toth2014, DemkowiczDobrzanski2015, Pezze2016, Degen2017}.  Recent experimental progress allows for a wide spectrum of applications of this form of enhanced metrology in magnetometry \cite{Budker2007, Taylor2008}, precision spectroscopy and frequency standards \cite{Wineland1992, Wineland1994} and the stabilization of atomic clocks \cite{Ludlow2015}. One of the most spectacular achievements is the quantum enhancement of the detectors that made the observation of gravitational waves possible \cite{Abbott2016}. By employing 
squeezed states of light \cite{Caves1981, Genoni2011, Schnabel2017} more sensitive phase measurements have been demonstrated \cite{Collaboration2011, Aasi2013}, even despite the inevitable noise \cite{Demkowicz2013}. The recently launched European Flagship in Quantum Technologies has quantum metrology and sensing as one of its pillars which clearly illustrates the potential scope of the approach of using quantum systems for enhancing measurement precision \cite{Dowling2015, Schleich2016}. \\
In this tutorial we will focus on one specific albeit relevant aspect of quantum metrology, namely the theoretical study of achievable precision bounds in frequency estimation with open quantum systems \cite{Huelga1997, Escher2011, DemkowiczDobrzanski2012}. This is a fundamental problem that underpins many situations of practical interest, ranging from precision spectroscopy to magnetic sensing \cite{Wineland1992, Balasubramanian2008, Zhao2012, Cai2013, Romach2015}. Our aim is to present in a concise form recent theoretical work analyzing the ultimate precision that can be achieved in the presence of different forms of noise and in the limit of a very large number of repetitions/number of probes (asymptotic scaling) \cite{Smirne2016, Haase2017}. Those provide theoretical lower bounds on the error of the estimation which can be saturated using some form of quantum resource, typically (albeit not necessarily) entangled input states \cite{Braun2017, Gorecka2017}.  We will show that the nature of the quantum evolution, which is influenced by unavoidable environmental fluctuations, has a direct impact on the achievable precision. In the noiseless scenario, the HL---corresponding to the $1/N^2$  scaling of the mean squared error with the probe number $N$---dictates the ultimate precision attainable in any problem in which the parameter is locally and unitarily encoded onto each of the probes. Although the HL should be redefined when allowing for non-local encodings \cite{Boixo2007, RivasLuis2012, Demkowicz2017} due to, e.g., non-linear Hamiltonians or correlations mediated also by decoherence \cite{Dorner2012, Jeske2014, Guo2016, Yousefjani2017}, it is the uncorrelated noise \cite{Huelga1997}, which typically forces the asymptotic precision to follow the SQL, i.e., the mean squared error to still scale as $1/N$. However, we will show that the SQL can be overcome even in the presence of uncorrelated noise and discuss the different forms of asymptotic scaling that arise depending on the noise geometry \cite{Chaves2013, Haase2017}. To further scrutinize how the achievable precision depends on the noise form, we consider a microscopic model that exemplifies how the sensitivity of the quantum probe is diminished by the effect of environmental fluctuations.\\
Our aim is to provide a reasonably self-contained analysis and with that scope in mind we have structured this tutorial as follows. Starting from a classical setup, section \ref{sec:motivation} presents a succinct discussion on fundamental concepts in estimation theory. We want to first illustrate how errors propagate when performing indirect measurements and secondly show how evaluating the achievable precision in this type of measurements always involves an optimization. Fundamental limits to the precision of estimation can then be obtained in a very general framework. We further illustrate how the use of quantum resources facilitates the saturation of the ultimate precision bounds allowed by quantum theory and provide a phenomenological description of a noisy frequency estimation to show that environmental decoherence has a direct impact on the problem at hand. Section \ref{sec:OQS} provides a primer on open system dynamics for the unfamiliar reader so that we can discuss in section \ref{sec:ultimateLimits} a completely general scenario to derive the ultimate precision bounds in frequency estimation in the presence of noise. Section \ref{sec:realisticBounds} complements this abstract approach by linking the noisy evolution to concrete microscopic models. For completion, some additional results beyond the independent noise model are presented in section \ref{sec:beyond_indep_noise}. We present a final summary of results in section \ref{sec:conclusion}. A caution note on bibliography is necessary at this point. Given the specific topic under consideration, the study of asymptotic precision bounds in local parameter estimation, we have unavoidably left out many interesting works in the broad field of quantum metrology that concern either experimental realizations with small number of particles or theoretical issues that do not refer to the asymptotic limit. For those we refer the reader to the appropriate literature \cite{Ludlow2015, DemkowiczDobrzanski2015, Taylor2016, Budker2007, Schnabel2010, Szczykulska2016, Toth2014, Degen2017, Ma2011, Schnabel2017, Pezze2016, Cronin2009, Dowling2008}.

\section{Fundamentals of estimation theory}
\label{sec:motivation}
Let us first present a simple estimation problem, which allows us to illustrate how most basic notions and tools we are going to discuss in the coming sections have their roots in classical statistics and, in particular, in the quantification of measurement errors. We want to exemplify this along the lines of the following example.

Imagine, one has $N$ identical coins, where a flip of an individual coin either gives heads with probability $p_h$ or tails with $p_t$. We stress that the only necessity to introduce these probabilities is our lack of knowledge of the exact initial conditions of the coin flip. The introduction of these probabilities is a way to describe the experiment statistically, while each flip and the subsequent observation of head or tail is completely deterministic and will depend on certain parameters of the coin which are too complicated to access, hence we resort to the much simpler quantification of the coin via its probabilities for heads and tails. To that end, one way to proceed is to flip each coin $\nu$ times to estimate $p_h$. Then, the probability variable $X$ describing the number of heads after $\nu N$ tosses is distributed according to binomial distribution
\begin{equation}
B(X=x\vert p_h,\nu N) = \left(\begin{array}{c} \nu N \\ x \end{array}\right)\, p_h^x \, (1-p_h)^{\nu N-x},
\end{equation}
with $x$ the number of heads found. Our best guess for $p_h$ is then obviously $p_h=x/(\nu N)$: we simply take the ratio between the number of observed heads and the total number of tosses. Indeed, $p_h$ coincides with the expectation value $\E{X/(\nu N)}$. However due to the finite quantities $N$ and $\nu$, our guess will carry an error.
A natural way to quantify this error is the variance
\begin{equation}
\var{p_h}=\var{\frac{X}{\nu N}}=\frac{p_h(1-p_h)}{\nu N}.
\label{eq:varP}
\end{equation}
Note that the variance is never equal to zero, besides the two special cases $p_h=0,1$, while the best strategy is to flip as many coins as often as possible. However, it is important to stress that the probabilities used to calculate the variance are not known as they themselves are the parameters to be estimated. As a consequence, the variance will never vanish in practice since the certain determination of the probabilities would require an infinite number of tosses.

To go a step further, we now assume that we are able to determine a parameter which changes the result of a coin toss, let's say its roundness $r$. Therefore, using the statistical model, it will change the probability of finding head and we assume that we know the deterministic dependence $p_h(r)$. An estimate of $r$ is then immediately given by the inverse function~\cite{Note1} $r(p_h)=p_h^{-1}(r)$ and we can use error propagation to find the variance on our estimate of $r$,
\begin{equation}
\var{r}=\frac{\var{p_h(r)}}{[\D p_h(r)/\D r]^2} = \frac{p_h(r)[1-p_h(r)]}{\nu N} \left[\frac{\D p_h(r)}{\D r}\right]^{-2}.
\label{eq:varr}
\end{equation}
In essence, it will turn out that this equation (or slight variations) is the working horse for estimation tasks we will deal with in the following. While we will perform generalizations to include the particularities of quantum mechanics in our theoretical descriptions, any quantification in sensing experiments can be linked to it \cite{Degen2017}.

Indeed we may compare the results above with an experiment utilizing quantum coins, i.e. two level systems (qubits). We replace the coins by $N$ qubits, each in the state $\ket{x_+}$, an eigenstate of the Pauli spin matrix $\sx$. Instead of a toss, we perform a unitary operation $U =\exp\left(-i\phi\sigma_z/2\right)$ and measure the survival probability $p_s$ of the state (note that this is conceptually a Ramsey experiment \cite{Ramsey1950}):
\begin{equation}
p_s =\abs{\bra{x_+}U\ket{x_+}}^2= \cos^2 \frac{\phi}{2}.
\end{equation}
Following the same ideas as above, we determine $p_s$ by the number of qubits found in $\ket{x_+}$ divided by $\nu N$, consequently $\var{p_s}$  and $\var{\phi}$ are immediately given by Eqs.~\eqref{eq:varP} and \eqref{eq:varr} respectively. 

We can directly use these relations to derive a well known bound in frequency estimation following \cite{Huelga1997}. Here the role of $\phi$ is taken by $\w t$ where $\w$ is the frequency to estimate (i.e. the role of $r$) while $t$ represents the time required to perform the unitary transformation. Because this time limits the number of repetitions, we also rephrase $\nu = T/t$ in terms of a total time $T$ that we have at our disposal to perform the measurement. Hence, we use Eq.~\eqref{eq:varr} and obtain
\begin{equation}
\var{\w} = \frac{1}{N \nu t^2} = \frac{1}{N t T}.
\label{eq:INTROsql}
\end{equation}
This equation manifests the so called \textit{shot noise limit} \cite{Itano1993} or \textit{standard quantum limit} (SQL).
While this term is used in the context of experiments involving quantum mechanics, its true origin lies, as we saw above, in the finite sample size of the underlying probability distribution. In other words, this effect is inevitable when dealing with randomly distributed data. Crucially, in quantum mechanics every experiment includes probability as an inherent feature. Thereby note that quantum mechanics is a probability theory itself, however, it is non-contextual \cite{Bell1966, Kochen1968}. With the recent developments in quantum technologies, promising sensors exploiting quantum mechanics have been put into the near future, explaining the rising interest in the field of quantum metrology.

\subsection{The frequency estimation protocol - analyzing a specific measurement procedure}
\label{sec:FEP_analyzingDevice}

\begin{figure}[t!]
\includegraphics[width=\columnwidth]{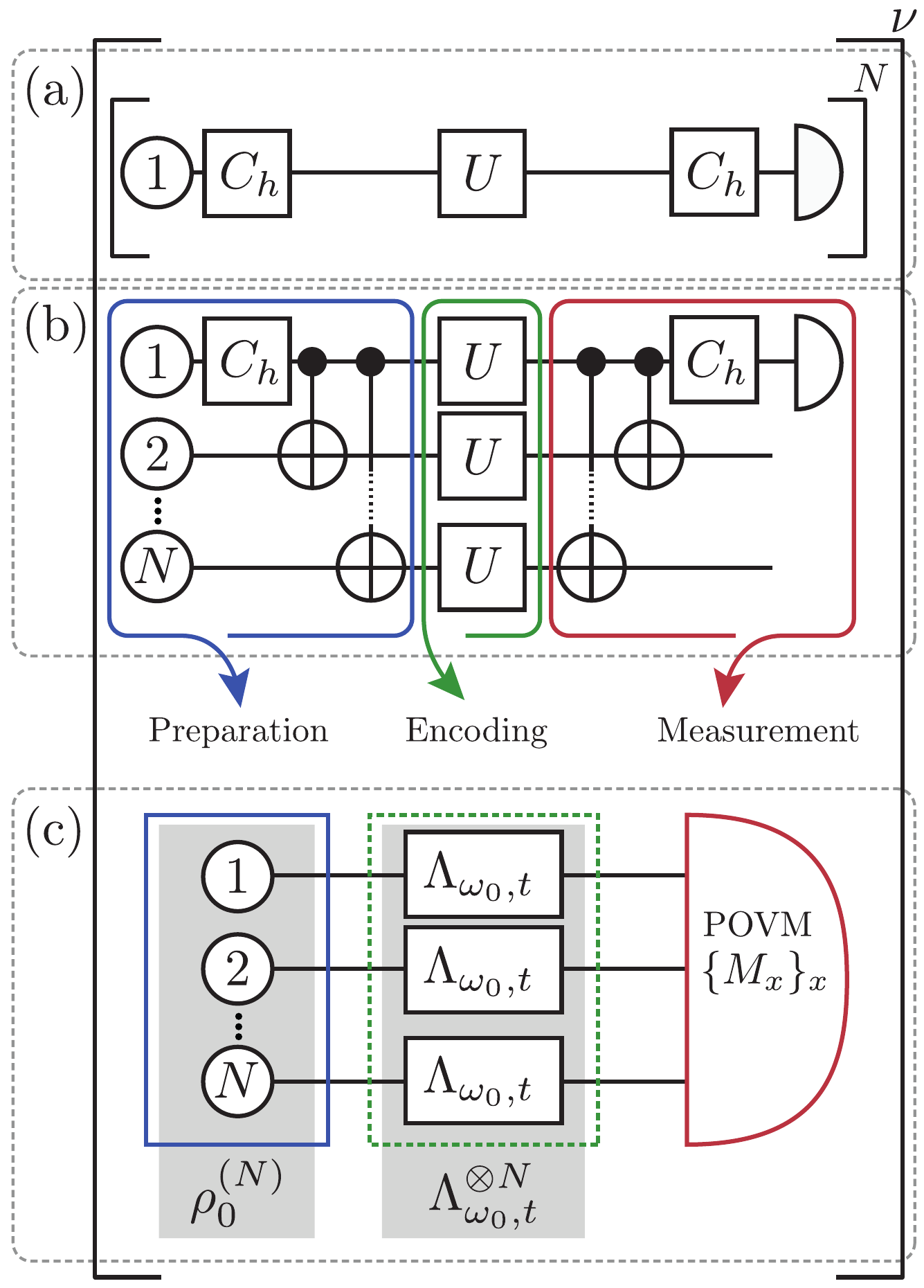}
\caption{\textbf{The frequency estimation protocol.} (a) Shows a single probe Ramsey type protocol for the quantum coins described in the main text. A Hadamard gate $C_h$ creates a state sensitive to the field during the free evolution $U$. After the second application of the Hadamard gate, the state is measured and the sequence is repeated. In (b), the $N$ probes are prepared in a GHZ state via the Hadamard and CNOT gates. The free evolution acts independently on each probe, which is the main characteristic of the FEP. Subsequently, the state is disentangled which allows to perform a measurement on the first probe only. Note that for other setups than the Ramsey scheme, the building blocks in the FEP may appear substantially different. All elements of (b) are assigned to either the preparation (blue), the encoding (green) or the measurement stage (red). These stages are generalized in the cFEP in panel (c). For the preparation, any state involving all $N$ probes is considered, while the product $\Lambda_{\w,t}^{(N)}=\Lambda_{\w,t}^{\otimes N}$ is allowed to describe any physical transformation, while the POVM invoked to describe the measurement has the only restriction to be independent of $\w$.}
\label{fig:freqEstProt}
\end{figure}

In this section, we want to analyze a specific measurement setup, in particular, we will use a Ramsey protocol \cite{Ramsey1950} that we utilize to measure the energy separation in a qubit. Indeed, the Ramsey experiment is nothing else than the quantum coins introduced in the section before. \\
Imagine we possess $N$ atoms and each can be modeled by two levels with a splitting of $\w$ (we take $\hbar=1$ throughout the whole work). For any of those atoms, we can assume the Hamiltonian $H_0 = \w \sz /2$. Following the Ramsey scheme outlined in Fig.\ref{fig:freqEstProt}(a), we initialize each qubit in its ground state $\ket{0}$ and apply a Hadamard gate $C_h$ which brings each of these qubits into an equally weighted superposition $\left(\ket{0}+\ket{1}\right)/\sqrt{2}$. Subsequently, these atoms evolve freely for a time $t$ during which they will collect a phase $\w t$ such that the state is given by $\left(\ket{0}+\mathrm{exp}(-i\w t)\ket{1}\right)/\sqrt{2}$. A second Hadamard gate will transfer the phase onto a population difference, which we measure via a suitable detector. The probability to find the qubit in $\ket{0}$ is then
\begin{equation}
p_{\w,t}\left(\ket{0}\right) = \abs{\bra{0}C_h e^{-it \w \sz/2} C_h \ket{0}}^2 = \cos^2\frac{\w t}{2}.
\label{eq:pOutRamsey}
\end{equation}
Indeed, we have $C_h\ket{0}=\ket{x_+}$ and thus everything is totally equivalent to the quantum coin example made in the introduction. However, the Ramsey experiment clearly illustrates the three stages present in a \textit{quantum frequency estimation protocol} (FEP) which will be the topic of the present tutorial. Other configurations are indeed possible \cite{Giovannetti2006, Giovannetti2011}. The first step is the \textit{preparation} of an input state for the \textit{probe}. Here, the probes are the qubits and the preparation is represented by the initialization and the application of the first Hadamard gate. It is followed by the \textit{encoding} which lasts for a \textit{time $t$}. The third step is then the \textit{measurement} of the probe, where we include the second Hadamard gate.  \\
Crucially, any specification of an achievable precision needs to be on a common ground. For that matter, we choose the number of probes $N$ and the total time $T$ as the \textit{resources} we have at our disposal. In particular, we assume the preparation and measurement process not to consume any resources, meaning the time needed for preparation and readout is negligibly small. An analysis relaxing this assumption can be found in \cite{Dooley2016}.

\subsection{The Cramer-Rao bound}
\label{sec:specificSetupTools}
As for the Ramsey setup, we will restrict to frequencies which are a linear parameter in the Hamiltonian $H_0$. Throughout this tutorial, $\w$ always denotes the parameter (or frequency) to be estimated. We already emphasize that in this context a \textit{probe} denotes the reduced quantum system we utilize for the estimation. This will become important later when noise is introduced into the system.

In general, a completed cycle of the FEP can be repeated several times. Obviously, the number of repetitions is fixed by the total time divided by the interrogation time, $\nu=T/t$. After each of these cycles, an outcome is detected. We collect all of these outcomes in the vector $\vec{x}$. To deduce $\w$ from the outcomes, an \textit{estimator} $\hat{\omega}(\vec{x})$ is constructed. Depending on the measured outcomes, the estimator yields an estimate $\hat{\omega}(\vec{x}) = \tilde{\omega}_0$ of the true value $\w$. Let's emphasize here that the estimator itself is a random variable, as the input (outcome of the measurement, i.e. the observations) is a random variable itself, i.e. $\vec{x}$ is one specific realization of $\vec{X}$. Therefore, it is possible to calculate different moments of the estimator, e.g., the expectation value  $\E{\bullet}_{\vec{x}}$ is taken with respect to the possible collections of outcomes $\vec{x}$, i.e.
\begin{equation}
\E{\bullet}_{\vec{x}} = \sum_{\vec{x}} p_{\w}(\vec{x}) \, \bullet.
\label{eq:pOutcomeCollection}
\end{equation}
The sum runs over all possible realizations of outcomes with $p_{\w}(\vec{x})$ being the probability that $\vec{x}$ is the realization obtained via the FEP. Note that we focus here and in the following on the case where we have a discrete set of possible outcomes, nonetheless, the whole description can be straightforwardly generalized to the case of a continuous set of outcomes. We adopt the notation $p_{\w}(\vec{x})$ for the conditional probability  $p(\vec{x}\vert\w)$ to obtain the set $\vec{x}$ given the parameter $\w$. However, after the data collection we can think of $p_{\w}(\vec{x})$ as the \textit{likelihood function} for $\w$ because the observations have already been made. Then, $p_{\w}(\vec{x})$ may be interpreted as a function of $\w$ quantifying how well different values would agree with the observed data set.

The explicit form of the estimator is not important for the further calculations, but we will always focus on estimators with the following properties \cite{Kay1993, Bos2007}.
\begin{itemize}
 \item  \textit{Unbiasedness}, which characterizes estimators that fulfill $\E{\hat{\omega}(\vec{x})}_{\vec{x}}= \w$. Conversely, an estimator is biased if $\E{\hat{\omega}(\vec{x})}_{\vec{x}}= \w + \beta$ where we have the bias $\beta \not = 0$.
 \item \textit{Consistency}, that is, for all $\nu>\nu^\prime$ there are $\epsilon(\nu^\prime),\delta(\nu^\prime)>0$ such that the probability $P(\abs{\tilde{\omega}_0-\w}<\epsilon) > 1 - \delta$. In other words, in the case of an infinitely large sample size, i.e. $\mathrm{dim}(\vec{x})=\xi\rightarrow\infty$, we have $\lim_{\xi\rightarrow\infty}\hat{\omega}(\vec{x})=\w$ and the estimator gives the true parameter.
\end{itemize}

\begin{figure}[t!]
\includegraphics[width=\columnwidth]{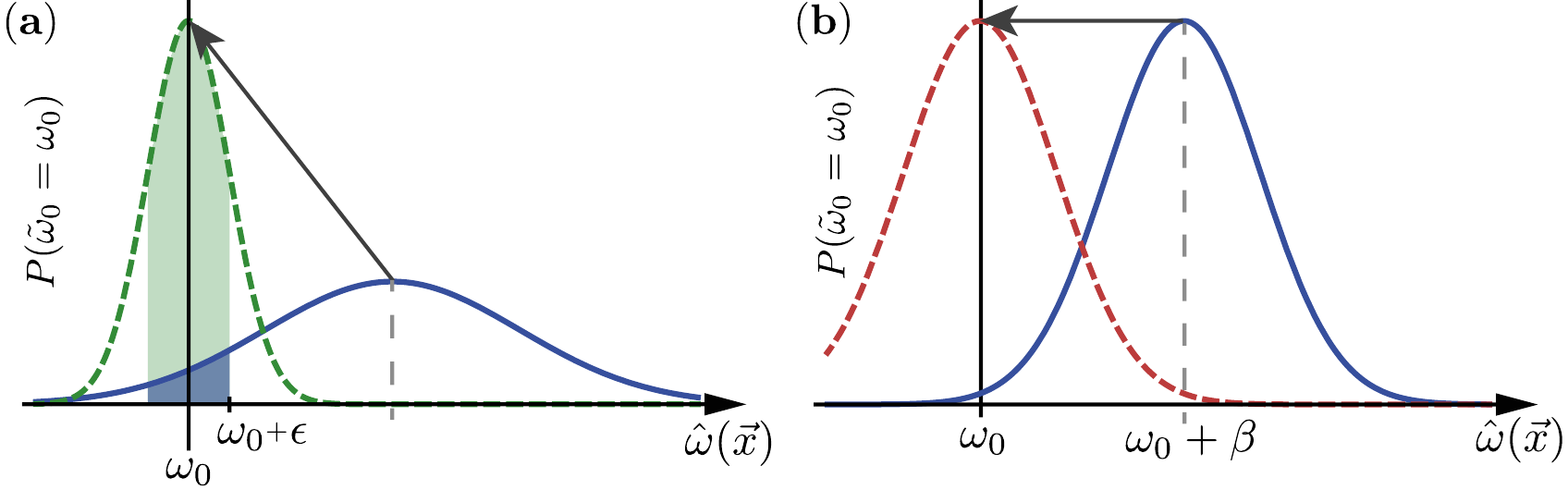}
\caption{\textbf{Properties of the estimator.} Both panels show the probability density (blue, solid) for an arbitrary estimator, for an arbitrary large data set $\vec{x}$. Likewise, the second curve represents the probability density for an enlarged data set. The estimator in (a) is consistent: further data collection removes any bias, while the probability to find $|\tilde{\omega}_0-\w|<\epsilon$ increases (shaded regions). Contrary, the estimator in panel (b) is only asymptotically unbiased, as the shape of the probability density does not change.}
\label{fig:estimatorProperties}
\end{figure}

Note that consistency implies \textit{asymptotic unbiasedness}, meaning that any bias $\beta$ vanishes for a large sample size. We stress that the converse is not true, see Fig.\ref{fig:estimatorProperties} for an illustration.

We define the precision of an estimator in terms of its \textit{mean squared error} $\Delta^2 \hat{\omega}$ (MSE),
\begin{equation}
\Delta^2 \hat{\omega} = \E{\left(\hat{\omega}(\vec{x})-\w\right)^2}_{\vec{x}}
\end{equation}
which is a natural choice as it measures the expected squared distance of the estimate $\tilde{\omega}_0$ from the true value $\w$. In particular, the MSE coincides with the variance of an unbiased estimator, defined as $\var{\hat{\omega}(\vec{x})} = \E{\left(\hat{\omega}(\vec{x})-\E{\hat{\omega}}_{\vec{x}}\right)^2}_{\vec{x}}$. While we will focus on unbiased estimators in the following, we will keep the notion of MSE instead of the variance.

For any unbiased estimator, its MSE can be bounded from below by the \textit{Cram{\'e}r-Rao bound} (CRB) \cite{Kay1993, Bos2007}
\begin{equation}
\Delta^2\hat{\omega}\geq \frac{1}{\nu \Fcl\left[p_{\w}\right]},\; \mathrm{where}\; \Fcl\left[p_{\w}\right] = \sum_{\vec{x}} \frac{[\partial p_{\w}(\vec{x})/\partial \w]^2}{p_{\w}(\vec{x})}
\label{eq:CRB}
\end{equation}
is the \textit{(classical) Fisher Information} (FI). Here, the sum runs over all possible collections of outcomes $\vec{x}$ and $p_{\w}$ is the same distribution as in Eq.~\eqref{eq:pOutcomeCollection}.
Any estimator achieving equality in the CRB is termed \textit{efficient}, but it is not a priori given that one can always find and an estimator of that kind \cite{Kay1993}.

The Fisher Information is a non-negative quantity which is additive for independent events \cite{DemkowiczDobrzanski2015}, i.e.
\begin{equation}
\Fcl[p_{\w}^{(1,2)}] = \Fcl[p_{\w}^{(1)}] \, + \, \Fcl[p_{\w}^{(2)}],
\label{eq:additivityFcl}
\end{equation}
where $p_{\w}^{(1,2)}(\vec{x}_1,\vec{x}_2)=p_{\w}^{(1)}(\vec{x}_1)p_{\w}^{(2)}(\vec{x}_2)$ is the joint probability distribution for the two events. This is of practical interest, as we will consider subsequent repetitions of the FEP which are uncorrelated by definition. Crucially, it is that additivity which is responsible for the $\nu$ in denominator of the CRB in Eq.~\eqref{eq:CRB}. Hence we are also able to give a precise meaning to $\vec{x}$ in the context of the FEP, which now contains the possible single run outcomes. Conversely, schemes employing an adaptive strategy, e.g. successively changing the measurement apparatus according to some prior acquired knowledge about $\w$, are not captured in the formulation by Eq.~\eqref{eq:CRB}, i.e. for them one cannot employ the sum given by Eq.~\eqref{eq:additivityFcl}. We will briefly discuss these strategies in Sec.~\ref{sec:beyond_indep_noise}.

Furthermore, the FI is a local quantity, as it only depends on the value of the probability distribution at the true value of the parameter and the first derivative. We may expand the probability distribution determining the observations,
\begin{equation}
p_{\w + \delta \omega}(\vec{x}) = p_{\w}(\vec{x}) + \frac{\partial p_{\w}(\vec{x})}{\partial \w} \delta \omega + O(\delta\omega^2),
\label{eq:FIlocality}
\end{equation}
where all terms fixing the FI are contained. Therefore, all distributions coinciding in zeroth and first order possess the same FI.

To exemplify the formalism now introduced, we directly calculate Eq.~\eqref{eq:varr} for the setup considered in the introduction, without passing through the Bernoulli distribution. We assume the FEP to be repeated $\nu$ times. After each cycle, the possible outcome of each qubit is either $\ket{0}$ or $\ket{1}$, hence $\vec{x}=(\ket{0},\, \ket{1})$. Thereby keep in mind that each qubit is independent from the other. Using the additive property we first obtain $\Fcl[\Pi_{n=1}^N p^{n}_{\w,t}] = N \Fcl[p_{\w,t}]$, then we note that $1-p_{\w,t}(\ket{0}) = p_{\w,t}(\ket{1})$ which  plugged into the CRB directly becomes Eq.~\eqref{eq:varr} with $p_h\mapsto p_{\w,t}(\ket{0})$. Furthermore, utilization of Eq.~\eqref{eq:pOutRamsey} directly yields the SQL Eq.~\eqref{eq:INTROsql}.

Crucially, as the CRB \eqref{eq:CRB} applies to any unbiased estimator, this proves that by choosing the number of heads (states $\ket{0}$) to be the random variable measured in $N$ rounds and simply from it inferring the parameter $\w$ by inverting Eq.~\eqref{eq:pOutRamsey}---so that the error propagation formula \eqref{eq:varr} applies---constituted the best strategy that could have been done.

In general, however, we have no guarantee that a given unbiased estimator $\hat{\omega}$ will be efficient---its MSE will saturate the CRB \eqref{eq:CRB} even in the asymptotic limit. Nevertheless, if $\hat{\omega}$ is built on a random variable (\emph{observable}) $O(\vec{x})$, which in turn relies on the outcomes $\vec{x}$ whose distribution is $\omega$-dependent, the error propagation formula \eqref{eq:varr} still applies and generally reads:
\begin{equation}
\Delta^2 \hat{\omega} = \var{O(\vec{x})}_{\w}
\,
\left[ \left.\frac{\partial \E{O(\vec{x})}_\omega}{\partial \omega} \right\vert_{\w}\right]^{-2},
\label{eq:errorPropagation}
\end{equation}
with $\left.\var{O(\vec{x})}\right]_{\omega}=\left\langle O(\vec{x})^2\right\rangle_{\omega}-\left\langle O(\vec{x})\right\rangle_{\omega}^2$ standing for the variance as before, where now $\left\langle O(\vec{x}) \right\rangle_{\omega}=\sum_{\vec{x}} p_\omega(\vec{x})\,O(\vec{x})$.

Nevertheless, let us note that one may always artificially construct $O_{\w}$ that yields an efficient $\hat\omega$ saturating Eq.~\eqref{eq:CRB}. Such an estimator, however, explicitly depends on $\w$, what makes the estimation procedure useless, unless one is only interested in sensing small parameter fluctuations around its known value. For a fixed $\w$ and outcome probability distribution $p_{\w}(x)$ (and its derivative $\dot{p}_{\w}(x)=\partial p_{\w}(x)/\partial\w $), one may always construct a (single-outcome) observable $O_{\w}(x)=\dot{p}_{\w}(x)/p_{\w}(x)$, which satisfies $\var{O_{\w}(x)}_{\w}
=\Fcl[p_{\w}]$ and $\partial\E{O_{\w}(x) }_\omega/\partial \omega|_{\w} = \Fcl[p_{\w}]$, so that at $\w$ Eq.~\eqref{eq:errorPropagation} indeed coincides with Eq.~\eqref{eq:CRB}.

For a further remark, performing a slight generalization of the CRB derivation and considering the estimation process of a smooth function of $\w$, one arrives at Eq.~\eqref{eq:CRB} that reads \cite{Bos2007}:
\begin{equation}
\Delta^2 \hat{g} \geq \frac{1}{\Fcl[p_{\w}]}\left( \frac{\partial g_{\w}}{\partial \w}\right)^2,
\label{eq:CRB_g}
\end{equation}
where $\hat{g}$ denotes now an unbiased estimator of some $g(\w)$. Importantly, the above expression proves that the efficient estimator saturating Eq.~\eqref{eq:CRB_g} can always be constructed from an efficient estimator of $\w$ by considering a smooth function $\hat{g}=g(\hat{\omega})$. Applying the error propagation to such an estimator with $O\equiv\hat{\omega}$ now in Eq.~\eqref{eq:errorPropagation}, we obtain
\begin{equation}
\Delta^2 \hat{g} = \Delta^2\hat{\omega}\; \left( \frac{\partial g_{\w}}{\partial \w}\right)^2, 
%\Leftrightarrow \Delta^2\hat{\omega} = \frac{\Delta^2 \hat{g}_{\w}}{\left(\partial g_{\w}/\partial \w\right)^2},
\end{equation}
which after substituting for the efficient $\Delta^2\hat{\omega}$ from Eq.~\eqref{eq:CRB} indeed yields the generalized CRB \eqref{eq:CRB_g}.

\subsection{Noiseless estimation with entangled states - obtaining Heisenberg limited precision}
\label{sec:HL}
In this section, we will show that quantum features can indeed improve the achievable precision below the SQL. Therefore, we use all $N$ probes together in a modified Ramsey setup. We stress again that all $N$ probes are absolutely equal and each local Hamiltonian is of the form given in terms of $\w\sz/2$. The modified setup, see Fig.~\ref{fig:freqEstProt}~(b), applies the Hadamard gate only on the first qubit, while there are controlled-not gates $C_{\mathrm{not}}^{(1,n)}$ on the $n$-th qubit, where the first qubit acts as the control. Instead of preparing the equally weighted superposition $N$ times, this arrangement creates an entangled \textit{GHZ}-state \cite{Greenberger1989} using the eigenstates of $\sz$,
\begin{equation}
\prod_{n=2}^{N}C_{\mathrm{not}}^{(1,n)} C_h^{(1)} \bigotimes_{m=1}^N \ket{0}_m = \frac{\ket{0}^{\otimes N}+\ket{1}^{\otimes N}}{\sqrt{2}} = \ket{\mathrm{GHZ}}.
\end{equation}
After the encoding, the gates are applied in the reverse order and the state of the first qubit is measured. The probability of finding it in $\ket{0}$ is $p_{\w,t}(\ket{0}) = \cos^2 (N\w t/2)$ and a direct combination with the CRB [or Eq.~\eqref{eq:varr}] yields (for an efficient estimator)
\begin{equation}
\Delta^2\hat{\omega}_{\mathrm{HL}} = \frac{1}{\nu t^2 N^2 } = \frac{1}{t T N^2 }.
\label{eq:HL}
\end{equation}
This scheme achieves a lower bound than the SQL, by an astounding factor of $1/N$, although we used the same number $N$ of probes and total time $T$ as before. This limit, scaling with $N^{-2}$, was named the \textit{Heisenberg Limit} (HL). It was argued to be the best achievable precision \cite{Bollinger1996} and indeed, this bound can be seen as an instance of the Heisenberg uncertainty relation \cite{Helstrom1976, Braunstein1996}. Later in Sec.~\ref{sec:QFI}, we will see how this connection can be made.

The role of entanglement in the preparation of the input state to obtain the HL has been extensively studied \cite{Giovannetti2004, Pezze2009, Demkowicz2014, Augusiak2016}. Indeed, the presence of entanglement is a strict requirement in the context of qubit probes in the FEP as considered here. However, we want to stress that it is \textit{not} true that entanglement is a necessary ingredient to beat the SQL or even achieve the HL in isolated quantum systems, when different estimation schemes are considered. We will comment on the latter in Sec.~\ref{sec:SubSQL_HO} and exemplify that a scaling similar to the HL can also be reached using a single probe repetitively.
\subsection{The impact of noise: Lindbladian dephasing}
\label{sec:LindbladDesphasing}
In a real world experiment, the evolution of the probes is unavoidably affected by noise. To give a flavor of the works presented later in this tutorial, we calculate an explicit example of an evolution under the influence of noise. Each probe is then an open quantum system \cite{Breuer2002, Rivas2012}, whose evolution is crucially shaped by the environment surrounding the probe. For the illustration of the peculiarities due to the presence of noise during the encoding stage, we restrict to a very specific kind of noise, i.e. we demand the noise to act independently but identically on each probe. Additionally, it has to be in the Lindblad form \cite{Rivas2014}. For simplicity, we restrict to pure dephasing, i.e. the probe's Hamiltonian commutes with the Hamiltonian introducing the noise, or in other words, in the basis which fixes $\sigma_z$, pure dephasing only damps the coherence elements of the probe's density matrix. This kind of noise can be seen, e.g., as a random fluctuation of the frequency, i.e., the parameter to be estimated. The evolution is then modeled by a master equation of Lindblad form,
\begin{equation}
\frac{\D \rho}{\D t} = -i [H_0,\rho] + \gamma \left( \sz \rho \sz - \rho \right),
\end{equation}
where $\gamma$ is a constant describing the decay strength of the noise.

We now repeat the calculations for the Ramsey scheme. For the scheme using $N$ probes in parallel, we mark all quantities with the subscript ``sep" (for separable), while the setup entangling the probes gets the subscript ``ent". We arrive at the probabilities
\begin{eqnarray}
p_{\w,t}^{\mathrm{sep}}(\ket{0}) &=& \frac{1+e^{-\gamma t}\cos\w t}{2}, \label{eq:pLindbladSep} \\ p_{\w,t}^\mathrm{ent}(\ket{0})  &=& \frac{1+e^{-N\gamma t}\cos N\w t}{2},
\end{eqnarray}
respectively, where we recognize the $N$ times higher oscillation frequency for the entangled state, however the exponential decay term stemming from noise obtains the same amplification. A subsequent calculation of the CRB yields
\begin{eqnarray}
\Delta^2\hat{\omega}^{\mathrm{sep}} &\geq&\frac{1-e^{-2\gamma t}\cos^2\w t}{N T t e^{-2\gamma t} \sin^2\w t}, \\
\Delta^2\hat{\omega}^{\mathrm{ent}} &\geq&\frac{1-e^{-2N\gamma t}\cos^2N\w t}{N^2 T t e^{-2N\gamma t} \sin^2N\w t}.
\label{eq:probsLindblad}
\end{eqnarray}
Indeed, these expressions are much more involved than the corresponding results for the noiseless cases, Eqs.~\eqref{eq:INTROsql} and \eqref{eq:HL}, and intuitively it is clear that the precision ought to possess an optimal interrogation time $t_\mathrm{opt}$. Note that for $t$ large enough, the derivatives of the probabilities in Eq.~\eqref{eq:probsLindblad} with respect to $\w$ vanish, which in turn causes the FI to vanish and hence the CRB diverges, see Eq.~\eqref{eq:CRB}. This is also the case for $t=0$, hence there has to be an optimal time of interrogation. To find this optimal point of operation, we minimize the CRB over the interrogation time, yielding
\begin{eqnarray}
t_{\mathrm{opt}}^{\mathrm{sep}} &=& \frac{k\pi}{2 \w} \overset{!}= \frac{1}{2\gamma}\; \Rightarrow \Delta^2\w^{\mathrm{sep}} \geq \frac{2\gamma e}{NT}, \label{eq:Lindblad_opt_Time}\\
t_{\mathrm{opt}}^\mathrm{ent} &=& \frac{k\pi}{2 n\w} \overset{!}= \frac{1}{2N\gamma}\; \Rightarrow \Delta^2\w^{\mathrm{ent}} \geq \frac{2\gamma e}{NT},\label{eq:Lindblad_opt_Time_ent}
\end{eqnarray}
where $k$ is an integer number. The achievable precision is exactly the same for both cases. This leads to the conclusion that product and entangled states (strategies) are metrologically equivalent under local dephasing Lindbladian noise. While this is certainly true for the scaling in the number of particles, a constant improvement of a factor $1/e$ can be achieved by using different entangled states (instead of GHZ) and measurement strategies \cite{Huelga1997, Escher2011, Kolodynski2013}.

At this point, let us stress a subtlety related with Eqs.~\eqref{eq:Lindblad_opt_Time} and \eqref{eq:Lindblad_opt_Time_ent} which involve a cyclic dependence of $\w$, $\gamma$ and the optimal time, while $\w$ is actually unknown. Importantly, these, and the following limits derived on $\Delta^2\hat{\omega}$ are always understood as the best possible precision achievable. One may always interpret these limits as a second step estimation process, where $\w$ is known roughly and the rest of the strategy is adapted according to the current knowledge. This may even be done via the choice of a suitable coordinate frame, see for example \cite{Haase2017}. For a more detailed discussion of this issue, see also Sec. \ref{sec:Saturation}.

\textit{Remark.---}
Note that the just derived bounds for a single probe ($N=1$) can be associated with the $T_2$ limit in quantum sensing \cite{Taylor2008, Degen2017}. Here it is used that $\gamma=1/T_2$, which results in an optimal time $t_\mathrm{opt}=T_2/2$, and therefore the precision is said to be $T_2$ limited.

%%%%%%%%%%%%%%%%%%%%%%%%%%%%%%%%%%%%%%%%%%%%%%%%%%%%%%%%%%%%%%%%%%%%%%%%%%%%%%%%%%%%%%%%%%%%%%%%%%%%

%%%%%%%%%%%%%%%%%%%%%%%%%%%%%%%%%%%%%%%%%%%%%%%%%%%%%%%%%%%%%%%%%%%%%%%%%%%%%%%%%%%%%%%%%%%%%%%%%%%%
%%%%%%%%%%%%%%%%%%%%%%%%%%%%%%%%%%%%%%%%%%%%%%%%%%%%%%%%%%%%%%%%%%%%%%%%%%%%%%%%%%%%%%%%%%%%%%%%%%%%
\section{A primer on open quantum system evolutions}
\label{sec:OQS}
This section is aimed at readers not familiar with the theory of open quantum systems, while experienced readers may skip this section as we will also introduce all required notation at its first appearance after this section. For a closer study of the topic, the reader is referred to the references \cite{Breuer2002, Nielsen2010, Rivas2012, Rivas2014, Breuer2016}.

The initial state $\rho_0$ of a closed system evolves according to the group of unitary operators $U(t)$ generated by the associated Hermitian Hamiltonian $H$, which for the sake of simplicity is to be assumed time independent for this section. The solution for the Schr{\"o}dinger equation of motion is then immediately given by
\begin{equation}
\rho(t) = U(t)\rho_0 U^\dagger (t) = e^{-itH} \rho_0 e^{iHt}.
\label{eq:oqs_unitary_evo}
\end{equation}
Crucially, this structure preserves the purity of the system, i.e. $\tr{\rho(t)^2}$ is conserved and equal to one for pure states. However, often the system of interest is in contact with an environment and hence the state of the now open system is obtained via the partial trace over the environmental degrees of freedom, i.e.
\begin{equation}
\rho(t) = \ptr{E}{U_{\mathrm{total}}(t)\,\rho_{0,\mathrm{total}} U_{\mathrm{total}}^\dagger(t) }.
\end{equation}
The total evolution operators $U_{\mathrm{total}}(t)$ are fixed by the form of the environment and the accompanied interaction with the open system. Usually the explicit form of these operators escapes our access due to the size of the environment or other technical restrictions which forbid its observation. Often, the structure of the specific environment isn't even known exactly and one employs a model which introduces the dynamics observed in experiments, e.g. the spin-boson model \cite{Leggett1987} as we will do later in this work. As a result of the partial trace, the evolution of the open system state is given by a Hamiltonian term $H^\prime(t)$, not necessarily equal to $H$, plus a dissipator $\mathcal{D}_t$, which captures the influence of the environment on the open system and assembles a so called quantum master equation,
\begin{equation}
\frac{\mathrm{d}}{\mathrm{d}t} \rho(t) = -i [\rho(t),H^\prime(t)] + \mathcal{D}_t\left[\rho(t)\right].
\end{equation}
Note that we already focused on the case where the equation is time-local, i.e. the evolution of $\rho(t)$ only depends on the current time point specified by $t$ and is independent of the previous history. This form can be obtained explicitly with the Born approximation (justified by the weak coupling between the open system and the environment) \cite{Breuer2002}, however, a derivation using the time-convolutionless technique also yields that result without invoking such an assumption.

A practical form of the master equation is the standard form, where the dissipator can be written as \cite{Rivas2012}
\begin{equation}
\mathcal{D}_t\left[\rho(t)\right] = \sum_{r=1}^{d^2-1} \gamma_r(t) \left[ V_r \rho(t) V_r^\dagger - \frac{1}{2} \lbrace V_r^\dagger V_r,\rho(t) \rbrace\right]
\end{equation}
with $d=\mathrm{dim}(\rho)$, possibly time dependent decay rates $\gamma_r(t)$, the operators $V_r$ and the anti-commutator $\lbrace A,B\rbrace = AB + BA$.

For an open system evolution, we define an analogous relation to Eq.~\eqref{eq:oqs_unitary_evo},
\begin{equation}
\rho(t) = \Lambda_{t\leftarrow 0}\left[ \rho_0 \right] = \sum_{r=1}^R K_r(t)\rho_0 K_r^\dagger(t),
\end{equation}
where $\Lambda_{t\leftarrow 0}$ is a so called dynamical map \cite{Rivas2012} or quantum channel \cite{Nielsen2010} evolving the state from time $0$ to $t$ and  $R$ is the rank of the evolution with $R\leq d^2$. For the second equality we used the Kraus-representation of $\Lambda_{t\leftarrow 0}$, invoking the Kraus operators $K_r(t)$ fulfilling $\sum_{r=1}^R K_r^\dagger(t) K_r(t)=\id$ which guarantees the preservation of the trace of $\rho$. It is important to stress that this representation only exists, iff the dynamical map $\Lambda_t$ is completely positive (CP) \cite{Nielsen2010}, i.e. $\left(\Lambda_{t_2\leftarrow t_1}\otimes \mathcal{I}\right) [\rho \otimes \id_{\mathrm{\dim \rho}}] \geq 0$ for any $\rho \geq 0$ and $\id_d$ the identity of dimension $d$ and $\mathcal{I}$ the identity map.

The standard form allows for an easy characterization of the induced dynamics of the open system via the decay rates $\gamma_r(t)$. First of all, if $\gamma_r(t)\geq 0\; \forall t,r$ the induced evolution is always completely positive \cite{Rivas2014} and the solution of the Lindblad equation can always be written via the Kraus operators. For rates $\gamma_r(t)<0$ the complete positivity has to be validated via the positivity of the Choi matrix \cite{Choi1975}.

The case when all rates $\gamma_r$ are constant implies that the dynamical maps form a semigroup whose elements only depend on the length of the evolved time intervall, $\Lambda_{t_2\leftarrow t_1}=\Lambda_{t_2-t_1}$, and the semigroup composition law is fulfilled \cite{Rivas2012, Rivas2014}, i.e.
\begin{equation}
\Lambda_{t_2+t_1} = \Lambda_{t_2} \circ \Lambda_{t_1}\quad \forall t_1,t_2.
\end{equation}
Such an evolution is called time-homogeneous (due to the constant rates) and was originally defined as the criteria for a Markovian evolution \cite{Rivas2014}. More recently, different definitions of non-Markovianity have been introduced \cite{Rivas2012, Breuer2016}, defined by either the violation of CP-divisibility \cite{Rivas2010} or the non-monotonicity of the trace distance \cite{Breuer2009}. The former states that any dynamical map which can be composed via
\begin{equation}
\Lambda_{t_3\leftarrow t_1} = \Lambda_{t_3\leftarrow t_2} \circ \Lambda_{t_2\leftarrow t_1}\quad \forall t_3 \geq t_2 \geq t_1,
\end{equation}
where $\Lambda_{t_3\leftarrow t_2}$ is a completely positive and trace preserving map, describes a Markovian evolution. The latter definition states that iff the evolution is Markovian, it holds that for any two states $\rho$ and $\sigma$
\begin{equation}
||\rho(t_2)-\sigma(t_2)||_1 \leq || \rho(t_1)-\sigma(t_1) ||_1 \quad \forall t_2 \geq t_1,
\end{equation}
where $||A||_1 = \tr{\sqrt{AA^\dagger}}$ is the trace norm. This notion of non-Markovianity is often associated with a backflow of information from the environment to the open system. We stress that these definitions are not equivalent. However, in both cases, the so called time-inhomogeneous evolution defined by always positive but time dependent rates in the master equation, is counted as Markovian.
Consequently, non-Markovianity corresponds to rates $\gamma_r(t)$ which are allowed to be negative, at least for some $t$. Crucially, this will violate both criteria. To this end we want to emphasize that the specific definition of non-Markovianity, apart from the semigroup composition law, does not play any role for the interpretation of the results presented in this tutorial.

\section{Ultimate Precision Limits - Analyzing arbitrary quantum channels, initial States and measurements}
\label{sec:ultimateLimits}

To evaluate the highest achievable precision of a measurement device operating in the quantum regime it is necessary to specify additional boundary conditions. At first, let us mention the possibility of different initial states which can be prepared. As we have already seen, the employment of entanglement yields a higher scaling of the achievable precision in the number of probes. Second, during the encoding period, the noise affects the system. While this may also be the case for a classical measurement device, here the noise can be purely quantum, e.g., a quantized radiation field \cite{Breuer2002}. And third, one can consider different possible measurement procedures. Realistically however, experimental realizations often limit this pool to a finite set. \\

Consequently, we consider a framework where these possibilities are taken into account. Therefore we generalize the FEP to the \textit{frequency estimation protocol for arbitrary quantum channels} (cFEP) within the \textit{independent noise model} and arbitrary initial states as well as arbitrary measurements. It is sketched in Fig.~\ref{fig:freqEstProt}~(c). In a first step, the $N$ probes are prepared in an arbitrary but chosen state. The specific properties of this state, i.e., whether it carries coherence or correlation, are transferred to an optimization involving all possible input states. Subsequently, the probes evolve for the encoding time $t$. The evolution of each single probe's reduced state is described via a completely positive and trace preserving (CPTP) \textit{dynamical map} \cite{Rivas2012} or equivalently a CPTP \textit{quantum channel} \cite{Nielsen2010}. We denote this channel by $\Lambda_{\w,t}$ which acts on the total input state $\rho_0^{(N)}$ of all $N$ probes as
\begin{equation}
\rho_{\w,t}^{(N)} = \Lambda^{\otimes N}_{\w,t}\left[\rho_0^{(N)}\right].
\end{equation}
The definition of the total map as the product of each single qubit channel, i.e. $\chan = \bigotimes_{n=1}^N \Lambda^{(n)}_{\w,t}$ is a necessity of the independent noise model. It ensures that all probes undergo the same evolution, i.e. the impact of noise on each probe is individual but identical, while it forbids  direct and environmentally mediated interactions of the probes during the interrogation time. The index $\w$ reminds that the channel possesses a dependence on the parameter to be estimated.

The last step in the protocol is the measurement. Again,  this is kept completely general in terms of the allowed measurements, i.e., they may be local on a single probe or global measurements on an arbitrary number of the probes. Needless to say, the choice of the measurement will fix the probability distribution of outcomes, which in turn fixes the Fisher Information and therefore the CRB. In this section, we will see how in the 	quantum framework it is possible to get an explicit form for the best possible precision, maximized over all the measurement procedures.
\subsection{Quantum Fisher Information and Quantum-CRB}
\label{sec:QFI}
Indeed, a chosen measurement immediately transfers a statistical operator (quantum state) to a (classical) probability distribution. A generic quantum measurement is described by a \textit{positive operator valued measure} (POVM), $\lbrace M_x \rbrace_x$ whose elements are positive-semidefinite operators associated with outcome $x$ for which it holds $\sum_x M_x = \id$. Choosing a POVM fixes the probability distribution $p^{(N)}_{\w,t}$, i.e. the probability to obtain outcome $x$ is $p^{(N)}_{\w,t}(x)=\tr{M_x\rho^{(N)}_{\w,t}}$. Following \cite{Helstrom1976, Braunstein1994}, the maximization of the FI over all POVMs yields the \textit{Quantum-Fisher-Information} (QFI)
\begin{equation}
F_Q\left[\R\right]:=\max_{\lbrace M_x \rbrace_x} F_{\mathrm{cl}}[p^{(N)}_{\w,t}(x)] = \tr{\R L_{\w,t}^2}.
\label{eq:QFI}
\end{equation}
Here, $L_{\w,t}$ is the \textit{symmetric logarithmic derivate} (SLD) of the state $\R$, which itself completely determines the QFI. Note that here we restrict to POVMs independent of $\w$, otherwise additional contributions appear \cite{Seveso2017}. The SLD is implicitly defined as
\begin{equation}
\frac{\partial \R}{\partial \w} = \frac{1}{2}\left( L_{\w,t} \R + \R L_{\w,t}\right),
\end{equation}
which is an instance of the Lyapunov equation \cite{Kitagawa1977} and merely states one of the core problems in quantum metrology. There exist an explicit solution to this equation, namely in the basis that diagonalizes $\R$, $L_{\w,t}$ can be expressed as
\begin{equation}
L_{\w,t} = \sum_{\lbrace j,k\vert p_{jj} + p_{kk} \not = 0\rbrace} \frac{2}{p_{jj}+p_{kk}} \bra{j}\frac{\partial \R}{\partial \w}\ket{k} \, \ketbra{j}{k},
\end{equation}
where $p_{jk} =  \bra{j}\R\ket{k}$. However, the involved diagonalization renders this problem numerically infeasible for systems of large dimension. If the state $\R$ is pure, i.e. $\R=\ketbra{\psi_{\w,t}}{\psi_{\w,t}}$ the QFI immediately reduces to (we suppress the index $\w,t$ for readability)
\begin{equation}
F_Q\left[\ket{\psi}\right] =  4\left( \braket{\partial_{\w} \psi}{\partial_{\w} \psi} - \abs{\braket{\psi}{\partial_{\w} \psi}}^2\right).
\label{eq:pureStateQFI}
\end{equation}
Using this equation, it is straightforward to calculate the QFI in case of a noiseless, i.e. a unitary evolution. Assuming we can write the encoding Hamiltonian in the form $H=\w H_{\mathrm{red}}$, with some suitable $\w$-independent Hermitian operator $H_{\mathrm{red}}$, the quantum channel is directly given by $\Lambda_{\w,t}[\bullet] = U\bullet U^\dagger$ with $U=\exp\left(-it\w H_{\mathrm{red}}\right)$ and one arrives at
\begin{equation}
F_Q[U\ket{\psi}] = 4 t^2 \Delta^2 \, H\big\vert_{\ket{\psi}}.
\label{eq:QFIpureState}
\end{equation}
Crucially, $\Delta^2 H\big\vert_{\ket{\psi}}$ is nothing else but the variance of the Hamiltonian generating the dynamics taken with respect to the initial state $\ket{\psi}$ \cite{Note2}. Note that for $H=\w H_{\mathrm{red}}$ the QFI is always independent of $\w$ itself \cite{Paris2009}.

We emphasize that the statistical operator $\R$ is the quantum state of all $N$ particles at once and may contain correlations between the different subsystems. This reduces the additivity of the QFI to the case of uncorrelated states, i.e. $F_Q[\rho_{\w,t}^{\otimes N}] = N F_Q[\rho_{\w,t}]$, since this is the only case where the measurements are indeed independent \cite{Kolodynski2013}. Analogously to the classical case, this could be thought of as either a parallel measurement on $N$ probes or an $N$ times repetition of the same measurement on a single probe.
Furthermore, the QFI is convex under incoherent mixtures of quantum states \cite{Fujiwara2001, Alipour2015}, i.e. for valid states $\rho, \sigma, \tau$ with $\rho=\lambda\, \sigma + (1-\lambda)\, \tau$ and $0\leq \lambda \leq 1$ we have
\begin{equation}
F_Q[\rho] \leq \lambda\, F_Q[\sigma] + (1-\lambda)\,F_Q[\tau].
\label{eq:QFIconvexity}
\end{equation}
Hence, any mixing of states cannot increase the QFI.

An equivalent definition of the QFI can be given in terms of the purification $\ket{\Psi_{\w,t}}$ of the state $\R$. By lifting the state into an Hilbert space extended by $\mathcal{H}_E$, the common state can expressed via the pure state vector $\ket{\Psi_{\w,t}}$, where $\R=\ptr{E}{\ketbra{\Psi_{\w,t}}{\Psi_{\w,t}}}$. Then, the QFI can be expressed as the minimum over these purifications \cite{Fujiwara2008, Escher2011, Kolodynski2013}
\begin{equation}
F_Q[\R] = 4\min_{\Psi_{\w,t}} \braket{\partial_{\w}\Psi_{\w,t}}{\partial_{\w}\Psi_{\w,t}}.
\label{eq:purificationQFI}
\end{equation}
Indeed, the crucial role of the QFI is due to the fact that it bounds the achievable precision for any possible measurement. Recalling the CRB in Eq.~\eqref{eq:CRB} and the definition of the QFI in Eq.~\eqref{eq:QFI}, we arrive in fact at the \nolinebreak{\textit{Quantum Cram{\'e}r-Rao Bound}} (QCRB), stating that the estimation error, minimized over any possible measurement for any initial state $\rho_0^{(N)}$ is lower bounded by
\begin{equation}
\Delta^2 \hat{\omega} \geq \min_{t \in [0,T]} \frac{t}{T\,F_Q\left\lbrace\chan \left[\rho^{(N)}_0\right]\right\rbrace}.
\label{eq:QCRB}
\end{equation}
Note that we explicitly mention the minimization to be performed over the interrogation time to obtain the optimal performance for the particular input state $\rho_0^{(N)}$.

To simplify the notation, from now on we will denote a derivation with respect to $\w$ with a simple overdot, i.e. $\partial_{\w}\bullet = \dot{\bullet}$.
\subsection{Achieving maximal precision - bounding the QFI}
For the aim of finding the maximal achievable precision for an arbitrary quantum channel, the maximization of the QFI with respect to the initial state is inevitable. While we already removed the necessity of specifying a measurement (POVM) in section \ref{sec:QFI}, here we will explore how the optimization of the QFI with respect to the input state can be performed efficiently. The only ``free" parameters left are then the encoding time $t$ and the quantum channel itself. In any case, the result of the input state optimization will indeed depend on the channel, hence it is meaningful to define the $channel-QFI$ (cQFI), which is the maximum FI at a time $t$ achievable when input state and readout are optimal. We define the cQFI as in \cite{DemkowiczDobrzanski2012, Kolodynski2013},
\begin{equation}
\mathcal{F}[\chan] := \max_{\rho_0^{(N)}} F_Q\left\lbrace\chan [\rho_0^{(N)}]\right\rbrace.
\label{eq:cQFI}
\end{equation}
The task of maximization quickly becomes involved, although due to the convexity of the QFI, Eq.~\eqref{eq:QFIconvexity}, the set of states over which the optimization in Eq.~\ref{eq:cQFI} is performed can be confined to pure states. For an increasing probe number, it is not a priori given that the optimal input state grows trivially with $N$, e.g., like the GHZ states in Sec.~\ref{sec:HL}, but non-trivial correlations may become important for some channels when $N$ is increased. Since the dimension of the state grows exponentially with $N$, numerical computation becomes infeasible even for small $N$ rendering the cQFI out of reach for examinations of an asymptotic scaling law. However, the cQFI can be bounded in terms of the Kraus operators representing the channel on the single probe level. Therefore, we will give an idea of the procedure in the single probe cQFI and state the result for the arbitrary $N$ case.

To avoid the calculation of the SLD, one utilizes the purification-based definition of the QFI, Eq.~\eqref{eq:purificationQFI}. The channel at any fixed time $t$, $\Lambda_{\w,t}(t)[\rho_0]$, can be regarded as a unitary evolution of the system in an extended Hilbert space, i.e., $\mathcal{H}_{\mathrm{ext}}=\mathcal{H}_S \otimes \mathcal{H}_E$, and subsequent tracing of this extension, using the Stinespring dilation theorem \cite{Nielsen2010}. Specifically we have
\begin{eqnarray}
\Lambda_{\w,t}[\rho_0] &=& \ptr{E}{U_{\w}(t)\, \rho_0 \otimes \rho_E\, U^\dagger_{\w}(t)} \notag \\
&=& \sum_{j}^R K_j (t,\w)\, \rho_0\, K_j^\dagger(t,\w)
\end{eqnarray}
where $K_j(\w,t)$ are the Kraus operators representing $\Lambda_{\w,t}$ and $\rho_E$ is a state of the extending subspace, which can always be assumed to be pure in terms of a purification performed on the extending subspace. Since the convexity of the QFI restricts $\rho_0$ to be pure, $\rho_0 \otimes \rho_E$ is pure and hence we can invoke Eq.~\eqref{eq:purificationQFI}. All purifications can then be reached by rotating the fixed $\rho_E$ with a unitary acting only on the extending subspace, $V_{\w}\rho_EV_{\w}^\dagger$. Note that these unitaries will in general depend on the frequency $\w$ itself. Thanks to the locality of the QFI, we are allowed to write this unitary in terms of a Hermitian matrix $h$ independent from $\w$, $V_{\w}=\mathrm{exp}\left(-ih \w\right)$. Note that after performing the partial trace using the rotated environmental state, the whole transformation boils down to a rotation of the channel's Kraus operators, i.e. we have $\tilde{K}_i(\w,t) = \sum_{j}^R (V_{\w})_{ij} K_j(\w,t)$ with $R$ the rank of the channel (note that this conversely implies that the dimension of $\mathcal{H}_E$ is $R$).
Taking the trace yields the cQFI as
\begin{eqnarray}
\mathcal{F}[\Lambda_{\w,t}] &=& 4 \max_{\rho_0} \min_{h} \tr{\sum_{i=1}^R \dot{\tilde{K}}_i(t,\w) \, \rho_0 \, \dot{\tilde{K}}_i^\dagger(t,\w)}, \notag \\
&&\mathrm{where} \notag \\
\dot{\tilde{K}}_i(t,\w) &=& \dot{K}_i(t,\w) - i \sum_{j=1}^R h_{ij} K_j(t,\w),
\end{eqnarray}
while higher order terms in $\w$ in $\dot{\tilde{K}}_i(t,\w)$ do not contribute due to the mentioned locality of the QFI.
We want to emphasize, that the environment used to employ Stinespring's theorem and, at the same time, the purification to obtain the cQFI is not necessarily a physical environment but merely a theoretical construct to avoid calculations involving the SLD.

The remaining maximization over (pure) input states $\rho_0$ is still a tedious task, especially for complicated channels or high dimensional systems. Importantly, the order of $\min$ and $\max$ cannot be exchanged. Nevertheless it turned out that an upper bound to the cQFI, based on the representation just calculated, can be efficiently determined, as it allows to exchange the order of the optimizations and hence the maximizations over input states can be performed. This approach has been named \textit{channel extension} and the idea is the following \cite{Fujiwara2008}: One extends the channel by an equally large Hilbert space, in particular, one assumes an arbitrary number of ancilla systems, which are not affected by the application of the quantum channel. However, if measurements on the new total space are considered, the information content measured by the cQFI can only grow, i.e., it is $\mathcal{F}[\Lambda_{\w,t}] \leq \mathcal{F}[\Lambda_{\w,t}\otimes \id]$. The total state of the system and the ancillas $\ket{\Psi_{\mathrm{SA}}}$ may be entangled, but can be assumed to be pure. After performing the partial trace over the (artificial) ancillas, one obtains [see Eq.~\eqref{eq:purificationQFI}]
\begin{eqnarray}
\mathcal{F}[\Lambda_{\w}(t)\otimes \id] &=& 4 \max_{\rho_\mathrm{S}} \min_{h} \ptr{\mathrm{S}}{\rho_\mathrm{S} \sum_{i=1}^R \dot{\tilde{K}}_i^\dagger(t,\w) \, \dot{\tilde{K}}_i(t,\w)} \notag \\
&=& 4 \min_{h} \left|\left|\, \sum_{i=1}^R \dot{\tilde{K}}_i^\dagger(t,\w) \, \dot{\tilde{K}}_i(t,\w)\, \right|\right|
\label{eq:channelExtendedcQFI}
\end{eqnarray}
with $||\bullet||$ the operator norm \cite{Note3}. In the second equality, we used that $\rho_\mathrm{S} = \ptr{A}{\ketbra{\Psi_{\mathrm{SA}}}{\Psi_{\mathrm{SA}}}}$ is now mixed and thus both optimization domains are convex. Hence we are able to exchange the order of $\min$ and $\max$  by virtue of the minmax theorem \cite{Rockafellar1970} and, subsequently, the maximum over the states can be calculated by means of the Cauchy-Schwarz inequality. In particular, $\max_{\rho_\mathrm{S}} \tr{\rho_s A} = \left|\left| A \right|\right|$ for any operator $A$ since $\rho_\mathrm{S}$ is positive with $\tr{\rho_\mathrm{S}}=1$.

In principle, for the case of $N$ probes building up the cFEP the single probe result could be derived directly, however, the problematic exponential increase of Hilbert space's dimension remains. Luckily, one can further bound the channel extended cQFI for $N$ probes in terms of the single channel Kraus operators. When the global channel for the common state of the probes is given by $\Lambda_{\w,t}^{\otimes N}$, it can be shown that \cite{Fujiwara2008, Kolodynski2013}
\begin{eqnarray}
\mathcal{F}[\Lambda_{\w,t}^{\otimes N}] &\leq &  \mathcal{F}[(\Lambda_{\w,t}\otimes \id)^{\otimes N}] \notag \\
&\leq & 4 N \min_{h(N)} \left[||\alpha_{\tilde{K}}|| + (N-1)\, ||\beta_{\tilde{K}}||^2 \right] \notag \\ &\equiv & \mathcal{F}^\uparrow[\Lambda_{\w,t}^{\otimes N}], \label{eq:upperBoundcQFI} \\
\text{with}  \notag \\
\alpha_{\tilde{K}} &=&  \sum_{i=1}^R \dot{\tilde{K}}_i^\dagger(t,\w) \, \dot{\tilde{K}}_i(t,\w), \notag \\
\beta_{\tilde{K}} &=&  i \sum_{i=1}^R \dot{\tilde{K}}_i^\dagger(t,\w) \, \tilde{K}_i(t,\w). \notag
\end{eqnarray}
We stress the dependence of the optimal $h$ on the number of probes, i.e. the minimization has to be performed for every $N$. It has been discussed, that indeed this bound provides useful estimates of the QFI for all $N$, even in the asymptotic regime $N\rightarrow\infty$ \cite{DemkowiczDobrzanski2012, Kolodynski2013} and, in fact, this will be the basis for the results presented in the next sections. Indeed the cQFI in Eq.~\eqref{eq:channelExtendedcQFI} and the bound in Eq.~\eqref{eq:upperBoundcQFI} coincide for $N=1$, as well as when considering any Kraus representation,  i.e., $h$, such that $||\beta_{\tilde{K}}||=0$ (in this case one can show \cite{Kolodynski2013} that the second inequality in Eq.~\eqref{eq:upperBoundcQFI} is saturated). In the latter case, it might still be convenient to consider Kraus representations such that $||\beta_{\tilde{K}}||\not=0$ and the optimal $h$ in Eq.~\eqref{eq:upperBoundcQFI} for each finite value of $N$. This provides the so-called \textit{finite-$N$ channel extension} method, which plays a crucial role in frequency estimation, in order to determine how the optimal evaluation time depends on $N$ and, hence, the best possible scaling of the precision obtained by optimizing also over $t$ \cite{Kolodynski2013, Smirne2016}. In any case, the bound requires intensive numerical effort, but can be cast into a semidefinite program to perform the minimization efficiently \cite{DemkowiczDobrzanski2012, Kolodynski2013}. Note that, besides this channel extension method, also other methods have been proposed and developed in the literature \cite{Escher2011, DemkowiczDobrzanski2012, Alipour2014}.
\subsection{Saturation of the (Quantum-)CRB}
\label{sec:Saturation}
Let us now discuss the attainability of the (Q)CRB. The first thing one has to keep in mind is that one is free to choose $t\ll T$ which increases the number of repetitions $\nu=T/t$. This provides more measurement data gathered over the total time $T$, and hence can lead to better precision which then improves at a classical rate $\sim 1/\nu$.

The chain of inequalities for the (Q)CRB mentioned so far is given by
\begin{equation}
\Delta^2\hat{\omega}\cdot T \overset{(1)}{\geq} \min_t \frac{t}{F_{\mathrm{cl}}} \overset{(2)}{\geq} \min_t \frac{t}{F_Q} \overset{(3)}{\geq} \min_t \frac{t}{\mathcal{F}}\overset{(4)}{\geq} \min_t \frac{t}{\mathcal{F}^\uparrow},
\label{eq:saturation_eqs}
\end{equation}
which has been bounded by several optimization procedures.
\\

\textit{Inequality (1). ---} In fact, one should keep in mind that the saturability of inequality (1) is a non-trivial issue and strongly depends on the properties of the estimator and therefore classical data processing \cite{Paris2009}. We already mentioned that any efficient estimator which is unbiased will achieve the CRB, however, such an estimator may not even exist \textit{globally}, i.e. for any arbitrary value of $\w$.

An often constructed estimator is the \textit{maximum likelihood estimator} (MLE) which profits from the collection of a large data set (Consider for example the estimator for a sample mean $\bar{x}=\sum_{i=1}^\nu x_i/\nu$). Specifically, it is the estimator maximizing the likelihood $p_{\w}(\vec{x})$ from Sec.~\ref{sec:specificSetupTools}. One can show that the asymptotic probability distribution of the MLE is a normal distribution with mean $\w$ and variance $F_\mathrm{cl}$, i.e. saturating the CRB \cite{Kay1993}. While always being consistent, one also has to keep on mind that an MLE may only be \textit{asymptotically unbiased}, i.e., the bias $\beta$ vanishes asymptotically for a large sample size.

On the other hand, in the regime of a finite data set, saturation is as mentioned not guaranteed in general. More specifically, this is true at least on a global level, i.e., irrespectively of the true value $\w$. In particular, a globally efficient estimator can only be found if the underlying probability density belongs the so called exponential family \cite{Kay1993}. Crucially, the normal distribution belongs to that family and hence the asymptotic saturability can be understood as an instance of the central-limit-theorem \cite{Kay1993, Guta2007}.

However, one can always follow a \textit{local approach} to saturate the CRB  locally at a point $\w = \omega_L$, where one constructs a \textit{locally unbiased estimator} which satisfies a local unbiasedness condition \cite{Fujiwara2006, DemkowiczDobrzanski2015},
\begin{equation}
\left.\frac{\partial}{\partial \w} \langle \hat{\omega}(\vec{x})\rangle_{\vec{x}}\,\right \vert_{\w=\omega_L} = 1.
\end{equation}
Indeed, when this condition is imposed during the derivation of the CRB, one exactly obtains Eq.~\eqref{eq:CRB} with the only restriction that it is only valid (i.e. equality can be reached) for an interval $\w = \omega_L \pm \delta \omega$, as one also restricts the FI to this interval, which on the other hand is a local quantity anyways, see Eq.~\eqref{eq:FIlocality}.
One is tempted to believe, that such a constraint renders the whole formalism impractical, as this restriction is very much present in nearly every estimation scheme since a globally unbiased estimator can almost never be constructed in a useful manner, or can't even be found for the problem at hand \cite{Kay1993}. The usefulness of the local approach traces back to the fact that one often possesses preliminary information about the parameter, such that the scheme becomes applicable to the measurement of small fluctuations in the parameter as it is done in atomic clocks \cite{Ludlow2015}, gravitational wave detectors \cite{Collaboration2011, Aasi2013} or as in a quantum sensing scenario named ``slope detection" \cite{Degen2017} employing for example nitrogen-vacency centers in diamond \cite{Taylor2008} for magnetometry. Furthermore, one can think of the protocol as a ``second step estimation", where one roughly determines the parameter first and applies the presented protocol for further refinement. Indeed, the local approach may be considered as the one giving the lowest bound hence its analysis may be regarded as the most optimistic one, therefore the derived limits can be considered fundamental. Additionally, the call for locality may be relaxed by allowing adaptive measurements, that is a sequence of MLE estimators based on locally unbiased estimates is consistent and asymptotically efficient \cite{Fujiwara2006}.

For an approach employing an estimation of the whole parameter range, one has to resort to Bayesian inference techniques to frequency estimation \cite{Macieszczak2014}. Here, one requires a new notion of the Heisenberg limit, which is now $\sim\pi^2/N^2$ \cite{Jarzyna2015} and can be saturated employing adaptive schemes as examined in \cite{Berry2009, Brivio2010}.
\\

\textit{Inequality (2). ---} The second inequality turns into an equality by choosing the POVM which maximizes the FI $F_{\mathrm{cl}}$. In particular, this POVM is given in terms of the projectors into the eigenbasis of the SLD operator \citep{Braunstein1994}, which in most cases turns out not to be a practical, realizable choice. In the specific case of a unitary evolution we know that Eq.~\eqref{eq:QFIpureState} holds, and then the optimal measurement (and input state!) are given by an equally weighted superposition of eigenstates belonging to the Hamiltonian $H$, as this state maximizes the variance \cite{Giovannetti2006}, compare Eq.~\eqref{eq:QFIpureState}. More precisely, for the Hamiltonian $H=\sum_{n=1}^{N}\w H_{\mathrm{red}}^{(n)}$ (where all $H_{\mathrm{red}}^{(n)}$ are identical), the SQL is achieved by the product state $\ket{\psi}^{\otimes N}$ where
\begin{eqnarray}
\ket{\psi}=\underset{\ket{\phi}}{\mathrm{argmax}} \, \Delta^2 H_\mathrm{red}\vert_{\ket{\phi}},
\end{eqnarray}
while the HL is achieved by
\begin{eqnarray}
\ket{\psi^{(N)}}=\frac{\ket{\mu_\mathrm{max}}^{\otimes N}+\ket{\mu_\mathrm{min}}^{\otimes N}}{\sqrt{2}},
\end{eqnarray}
with $\ket{\mu_{\mathrm{max/min}}}$ the eigenvectors belonging to the maximal (minimal) eigenvalues of $H_\mathrm{red}$. These states are trivially also the ones maximizing the cQFI and are therefore able to saturate all bounds given in Eq.~\eqref{eq:saturation_eqs}.
\\

\textit{Inequality (3). ---} The saturation of the bound given by the cQFI $\mathcal{F}$ is given if all the conditions set by the maximization procedures of the FI are fulfilled. This requires the knowledge of an optimal input state for the QFI. In the case of a unitary evolution, these can be found by maximizing the variance of the Hamiltonian, as mentioned in the previous paragraph. For a general open-system dynamics, i.e. an arbitrary quantum channel, the state maximizing the cQFI cannot be generally found explicitly. However, note that if one finds a state and a measurement procedure such that $t/F_{\mathrm{cl}}$ has the same scaling as $t/\mathcal{F}$, one can also argue that the optimal strategy will have such a scaling, as long as the classical CRB (1) is saturated as well.
\\

\textit{Inequality (4). ---} This inequality may never be saturated, as the cQFI $\mathcal{F}$ is an upper bound on the QFI $F_Q$ itself. However, its scaling with the probe number may be reached asymptotically ($N\rightarrow\infty$), apart from a possible constant. Analogously to the arguing for inequality (3), if one finds a state and a measurement procedure such that $t/F_\mathrm{cl}$ has the same scaling as $t/\mathcal{F}^\uparrow$, given the saturation of the classical CRB (1), the optimal  strategy will possess the same scaling.

%%%%%%%%%%%%%%%%%%%%%%%%%%%%%%%%%%%%%%%%%%%%%%%%%%%%%%%%%%%%%%%%%%%%%%%%%%%%%%%%%%%%%%%%%%%%%%%%%%%%

\section{Realistic bounds on the precision}
\label{sec:realisticBounds}
\begin{figure}[t!]
\includegraphics[width=\columnwidth]{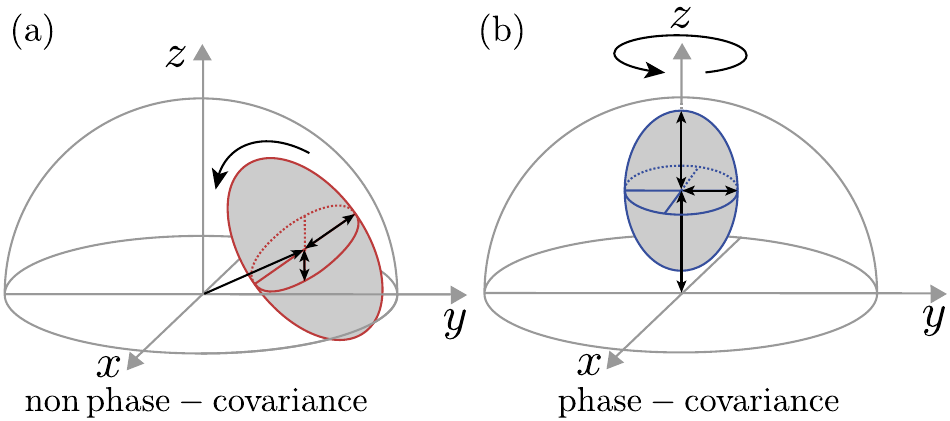}
\caption{\textbf{Geometrical picture of NPC and PC channels.} In panel (a) and (b) the grey shaded volume represents the available part of the Bloch sphere after the application of an arbitrary quantum channel, i.e. every valid Bloch vector has to point to a state within this volume. While for the NPC case in (a) the volume can take any shape and position, for a PC channel the volume has to be distributed around the $z$ axis such that it always possesses a rotational symmetry around that axis. This limits the allowed transformations of the volume to the ones mentioned in the main text. Intuitively, a rotation by an arbitrary angle around the $z$ axis at any point, before or after the application of the channel will change the picture in (a) while it does not in (b).}
\label{fig:PCspheres}
\end{figure}
The cFEP has been under heavy investigation to determine the best precision achievable under different circumstances and, in particular, the different kinds of noise during the encoding time. The main question is whether the ultimate limit is given by the SQL, or how close one can reach the HL. Thereby one has to keep in mind that these limits have to be understood in an \textit{asymptotic sense}, i.e. the number of available probes $N$ is large and tends to infinity. As we will see in this section, for a finite number of probes these asymptotic scalings may not yet hold and are usually worse.

Throughout this section, we will use the model from Ref.\cite{Haase2017} as a reference for realistic noise. It is capable of reproducing all the scalings we are going to present here, provided the parameters are chosen as such that the correct approximations to reach these regimes are justified. We model each probe via a qubit, which interacts with an infinite number of harmonic oscillators which in turn are independent from each other. This is the regularly invoked spin boson model for quantum dissipation \cite{Leggett1987}, where the Hamiltonian is specified as
\begin{eqnarray}
H &=& \frac{\w \sz}{2} + \sum_n \omega_n a^\dagger_n a_n \notag \\
&&+\left(\cos\Cang \frac{\sx}{2}+\sin\Cang \frac{\sz}{2}\right)\otimes \sum_n \left(g_n a_n+ g^*_n a_n^\dagger \right).
\label{eq:Hamiltonian}
\end{eqnarray}
The transition frequency of the qubit, $\w$, represents the parameter to be estimated. For the bath, we invoke the operators $a_n$ and $a^{\dagger}_n$, which are the bosonic annihilation and creation operators corresponding to mode $n$ with frequency $\omega_n$. The second line contains the coupling part of the Hamiltonian. Each environmental mode is coupled to the two-level system with strength $g_n$, while the parameter $\Cang$ defines the coupling angle between the $x$-axis and the direction of the coupling operator. This allows the identification of different scenarios: For $\Cang=\pi/2$ we have pure dephasing (or parallel noise with respect to $z$-direction) interaction, while for $\Cang=0$ we observe a purely transversal (or perpendicular) interaction. Under a weak coupling assumption between the qubit and the bath oscillators, one employs the second order time-convolutionless master equation for the reduced density matrix of the qubit alone \cite{Breuer2001, Breuer2002} and obtains
\begin{eqnarray}
\frac{\mathrm{d}\rho(t)}{\mathrm{d}t}= -i\left[\frac{\w}{2} \sz ,\rho(t)\right] + \gamma(t)\left(\bar{\sigma} \rho(t) \bar{\sigma}^\dagger -\rho(t)\right),
\label{eq:spinBosonME}
\end{eqnarray}
where we have abbreviated the quantities
\begin{eqnarray}
  \gamma(t) &=& \frac{\lambda}{\beta} \arctan\left(\omega_c t\right) \nonumber\\
\bar{\sigma} &=& \cos\Cang \sx+ \sin\Cang \sz.
\label{eq:ohmf}
\end{eqnarray}
In particular, the following assumptions are included in the derivation of Eq.~\eqref{eq:spinBosonME}: First, the spectral density used to describe the continuum of bath modes has the Ohmic form $J(\omega)=\sum_n g_n^2 \delta(\omega-\omega_n)\rightarrow\lambda \omega \exp\left(-\omega/\omega_c\right)$, where ``$\rightarrow$" describes the continuum limit. Here, $\lambda$ defines an overall coupling strength and $\omega_c$ is a cut off frequency, much larger than $\w$. Second, the total initial state of the bath is a thermal state with a low inverse temperature $\beta$. We want to stress further, that in general the direction of the noise in the master equation (here fixed by $\Cang$ in $\bar{\sigma}$) is a direct consequence of the direction fixed by the interaction Hamiltonian, but the preservation of the same functional dependence is a special case of the regime considered.

Performing the secular approximation \cite{Breuer2001, Breuer2002, Maniscalco2004, Fleming2010} during the derivation of the master equation corresponds to neglecting fast oscillating terms modifying the evolution \cite{Note4}. This is always justified in the scenarios where the free dynamics is much faster than the dissipative one. In particular, one separates the timescale $\tau_0 \sim \w^{-1}$ from the relaxation time of the system, $\tau_R$. As long as $\tau_0 \ll \tau_R$, terms oscillating with $\w$ are averaged out, which in turn decouples the populations and coherences of the qubit. Starting from Eq.\eqref{eq:spinBosonME}, one is left with the master equation
\begin{eqnarray}
\frac{\mathrm{d}\rho(t)}{\mathrm{d}t}&=& -i\left[\frac{\w}{2} \sz,\rho(t)\right] \nonumber \\&&+\gamma(t)\sum_{j=\pm,z} d_{j}\left(\sigma_j \rho(t) \sigma_j^\dagger -\frac{1}{2}\left\lbrace\sigma_j^\dagger \sigma_j,\rho(t)\right\rbrace\right)
\label{eq:masterEquationohmht}
\end{eqnarray}
with $d_+ = d_- = \cos^2\Cang$, while $d_z = \sin^2\Cang$.

We stress that both these master equations are CPTP, due to the positivity of $\gamma(t)\, \forall t\geq0$ \cite{Rivas2014}, which also categorizes them as time-inhomogeneous Markovian. Note that this is a consequence of the specific choice of the Ohmic spectral density, while no Markov-approximation has been performed. We will comment on the role of non-Markovianity later in this section. Furthermore, this model allows for a natural transition to master equations which are indeed Lindblad equations (i.e. their solution is a dynamical semigroup \cite{Breuer2002, Rivas2012}), where the decay rate $\gamma(t)$ is replaced by a constant. This can be achieved by taking the limit $\wc \rightarrow \infty$ \cite{Haase2017}, which corresponds to an infinite narrowing of the bath correlation functions (which decay as $\sim \omega_C^{-1}$), which is basically the necessary condition for the Markov approximation to hold.

As already explained in Sec. \ref{sec:Saturation}, the attainability of the QCRB employing the cQFI $\mathcal{F}$ as a lower bound to the achievable precision can be shown (at least up to a constant factor) by evaluating the precision, as quantified by the FI, for a specific measurement and initial state. For the cases taken into account here, it is enough to consider GHZ-states as the input and the parity operator, $P_x = \bigotimes_{n=1}^N \sx^{(n)}$ \cite{Ma2011} as the subsequent measurement. Using error propagation, Eq.~\eqref{eq:errorPropagation}, the error can be written as
\begin{eqnarray}
\Delta^2\hat{\omega}_{P} \cdot T =
t \, \frac{1-\langle P_x(t)\rangle^2}{\left| \langle \dot{P}_x(t)\rangle \right|^2}
\label{eq:parityPrecision}
\end{eqnarray}
where further calculations can be found in \cite{Brask2015, Haase2017}. Simulating that bound, provides us with a chain of inequalities
\begin{eqnarray}
\Delta^2\hat{\omega}_P\cdot T \geq \Delta^2\hat{\omega} \cdot T \geq \frac{t}{\mathcal{F}[\Lambda_{\w,t}^{\otimes N}]} \Rightarrow \Delta^2\hat{\omega} \cdot T \sim \frac{1}{N^\kappa},
\end{eqnarray}
where we justify the implication from the fact that when both sides approach 0 in the limit $N\rightarrow \infty$ as $N^{-\kappa}$, the same will be true for $\Delta^2\hat{\omega} \cdot T$.

\textit{Remark.---} It is important to keep in mind that in general the decay rates contained in the dissipator depend on the frequency $\omega$ to be estimated. Intuitively, this can be understood as a different part of the environmental spectral density is probed. These contributions are usually neglected, although they can change the magnitude of the QFI \cite{Haase2017} and they have also been considered in the context of Gaussian noise \cite{Szaifmmode-nelse-nfikowski2014}. However, in the model chosen in this work, this dependence is naturally removed by the choice of the Ohmic spectrum combined with the large cutoff frequency.

\subsection{The Zeno-Limit under phase-covariant noise}
\label{sec:ZL_PCnoise}
Let us start with noise which can be described using the master equation in Eq.~\eqref{eq:masterEquationohmht}. The secular approximation ensures that the noise induced during the evolution is \textit{phase-covariant} (PC) \cite{Holevo1993, Holevo1996, Vacchini2010, Lostaglio2017}. This requirement is defined through the condition that the channel generating the evolution commutes with any rotation $R_z[\bullet]=\exp\left(-i\phi \sz\right) \bullet \exp\left(i\phi \sz\right)$ of the qubit's state around the $z$ axis, i.e.,
\begin{equation}
\left[\Lambda_{\w,t},R_z\right] = \Lambda_{\w,t}\circ R_z - R_z \circ \Lambda_{\w,t} = 0
\label{eq:PC_condition}
\end{equation}
for any arbitrary angle $\phi$. In other words, the free evolution and the action of the noise commute. When the qubit's state is described in terms of its Bloch vector, phase-covariance results in a particular geometry of the available transformations made through the channel. The volume of available states always contracts isotropically in $x$ and $y$ direction, preserving the rotational symmetry around the $z$ axis. Furthermore, contractions and shifts along the $z$-axis are allowed, compare also Fig.\ref{fig:PCspheres}. Indeed, one can show that any dynamics fulfilling these conditions possesses a generator of the form
\begin{eqnarray}
\frac{\mathrm{d}\rho(t)}{\mathrm{d}t}&=& -i\left[\frac{\w + h_0(t)}{2} \sz,\rho(t)\right] \nonumber \\&&+\sum_{j=\pm,z} \gamma_j(t)\left(\sigma_j \rho(t) \sigma_j^\dagger -\frac{1}{2}\left\lbrace\sigma_j^\dagger \sigma_j,\rho(t)\right\rbrace\right)
\label{eq:PC_ME}
\end{eqnarray}
with suitable rates $\gamma_j(t)$ and a possibly time dependent Lamb-shift $h_0(t)$ \cite{Smirne2016}.

It has been shown that for any FEP where the channel can be described according to Eq.~\eqref{eq:PC_ME}, i.e., the channel is phase-covariant, the ultimate precision is always bounded from below \cite{Smirne2016} by the asymptotic scaling
\begin{equation}
\Delta^2\hat{\omega}_{\mathrm{Zeno,PC}} \cdot T  \;\geq\; \frac{C}{N^{3/2}},
\label{eq:scaling_ZL}
\end{equation}
for a suitable ($N$-independent) constant $C$. Furthermore, it was shown that such a limit can always be achieved (at most up to the constant factor) by means of a GHZ state.
First encounters with this scaling have been presented in \cite{Matsuzaki2011, Chin2012}, where it has been linked to the quadratic decay of transition probabilities in the environment on short time scales and was hence called the \textit{Zeno-Limit}. A general derivation for the case of pure dephasing can also be found in \cite{Macieszczak2015}. Indeed, as shown in \cite{Smirne2016}, the Zeno scaling emerges for all evolutions whose dynamics deviate from a Lindbladian (semigroup \cite{Rivas2012}) evolution at short time scales, while the precision collapses immediately to the SQL when the rates $\gamma_j(t)$ in Eq.~\eqref{eq:PC_ME} are replaced by constants \cite{Huelga1997}. In particular, the optimal interrogation time has been proved to scale as
\begin{equation}
t_{\mathrm{opt}}^{\mathrm{Zeno,PC}} \propto \frac{1}{N^{1/2}},
\end{equation}
for any evolution (apart from the unrealistic case of a full revival), thus showing explicitly how the optimal estimation strategy relies on measurements on shorter and shorter time scales.

\begin{figure}[t!]
\includegraphics[width=\columnwidth]{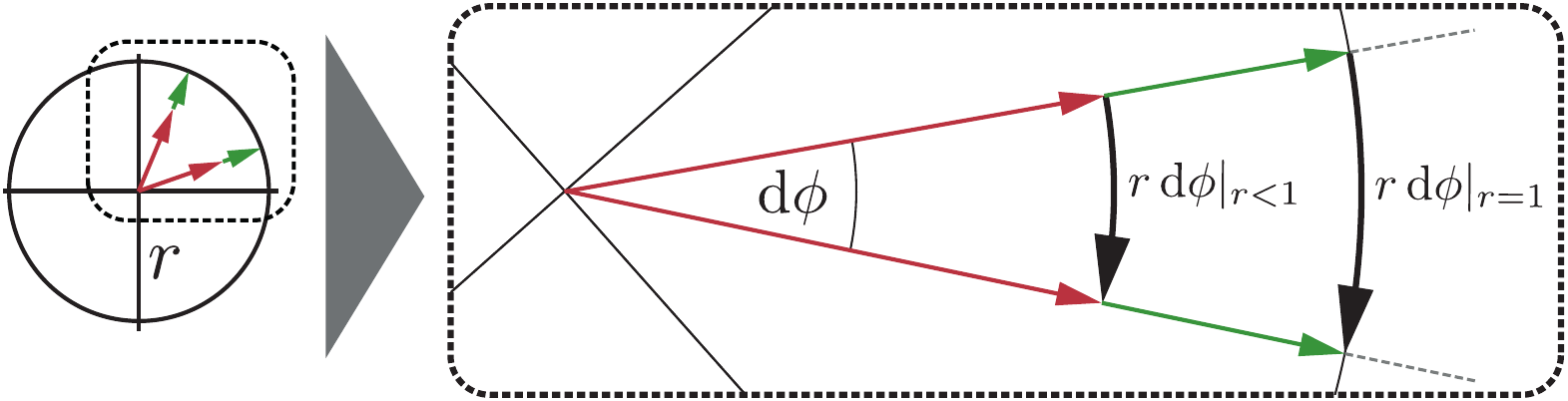}
\caption{\textbf{Geometric picture for the precision for a single probe ($N=1$).} The distance of two quantum states, i.e., the distance between two neighboring probability distributions can be visualized by two Bloch vectors of same length who only differ in a small phase angle $\mathrm{d}\phi$. Note that rigorously one should understand these as the projection of some Bloch vectors into the $xy$ plane. Decoherence processes decrease the length of these vectors (and their projections), hence the states are approaching each other, see the transition from green to red.}
\label{fig:geometrics}
\end{figure}

Note that a Lindbladian (semigroup) evolution corresponding to constant rates in Eq.~\eqref{eq:PC_ME} is generally an approximation to the real dynamics, as it relies on a coarse grained time resolution \cite{Breuer2002}, which neglects times where the environmental correlation functions aren't decayed yet. However, the total evolution of the system and the bath is always governed by a unitary evolution of a possibly time dependent Hamiltonian $H(t)$. Hence given an initial pure state of the system $\ket{\psi_S}$ and the total state $\ket{\psi}=\ket{\psi_S}\otimes\ket{\psi_E}$, the short time survival probability of the reduced state can be written as \cite{Smirne2016}
\begin{equation}
\bra{\psi_S}\Lambda_t\left[\ketbra{\psi_S}{\psi_S}\right]\ket{\psi_S} = 1-\alpha_S t^2 + O(t^3),
\end{equation}
which is always of the order $O(t^2)$ and $\alpha_S = \bra{\psi}H(0)^2\ket{\psi}-\bra{\psi_S}\ptr{E}{H(0)\ketbra{\psi}{\psi}H(0)}\ket{\psi_S}$. Hence we can understand a dynamics which is accurately described by a Lindblad master equation as a type of dynamics where the ``Zeno regime" (i.e. the regime where terms quadratic in time are relevant) is not accessible.
Moreover, since the information about $\w$ is encoded in the phase of the qubit's state, for the special single probe case (i.e. $N=1$) it was possible to show that the length of the Bloch vector's projection into the $xy$ plane determines the achievable precision. In this respect one may observe geometrically the balance between a long evolution time and the decoherence processes diminishing the achievable precision, for that compare also Fig.~\ref{fig:geometrics}.
Both affect the distance between the projections of the two states $\rho_{\w ,t}$ and $\rho_{\w +\mathrm{d}\w, t}$, which is given by the line element $r(t)\,\mathrm{d}\phi(t)=r(t)\,t\, \mathrm{d}\w$ where $r(t)$ is the length of the projection and $\mathrm{d}\phi(t)=t\,\mathrm{d}\w$ the phase difference. Obviously, the function $r(t)$ and $t$ counteract each other. While $t$ increases the phase difference and hence provides as better distinguishability of the states, $r(t)$ pulls the projections towards the origin and thereby decreasing the precision.

We want to emphasize again, that this limit is asymptotic, i.e., it is reached for a larger number of probes which in turn shifts the optimal interrogation time into the short time regime. This shifting can be motivated by the fact that entangled states do not only share their phase evolution, but also collectively gather fluctuations induced by the noise. Hence, the noise is ``naively $\sim N$-times stronger", i.e., the phase evolution is lost quicker. In the short time regime, the only time order left to contribute is the second one as shown above, which then yields the mentioned scaling.

As an example, we show here that the microscopic model given by the Hamiltonian \eqref{eq:Hamiltonian} induces the Zeno scaling, when we choose dephasing noise that is parallel to the signal encoding, i.e., $\Cang=\pi/2$. For a Ramsey measurement, we can calculate the CRB analogously to \cite{Chin2012}. We remark that for that choice of $\Cang$, Eq.\eqref{eq:PC_ME} and Eq.\eqref{eq:masterEquationohmht} coincide since the case of pure dephasing is always PC. Employing GHZ states (compare also Sec.\ref{sec:LindbladDesphasing}) we determine the survival probability
\begin{eqnarray}
p_{\w,t} = \frac{1}{2}\Bigg\lbrace 1 &+&\exp \left[-\frac{N\lambda}{\beta} \left(t \arctan(t\omega_c)-\frac{\log(1+t^2\omega_c^2)}{2\omega_c}\right)\right] \nonumber\\
&\times & \,\cos\left(Nt\w\right)\Bigg\rbrace.
\label{eq:transition_prob_PC_case}
\end{eqnarray}
Since the short time expansion yields $p_{\w,t} \approx 1- t^2 (N^2\w^2 - \lambda N \wc /\beta)/4+O(t^3)$, we expect the precision to be bound by the Zeno limit. Calculating the CRB employing the survival probability in Eq.~\eqref{eq:transition_prob_PC_case}, the subsequent derivations of the CRB with respect to $\w$ and $t$ yield the optimality conditions,
\begin{eqnarray}
\topt &=& \frac{k\pi}{2N\w} \; \mathrm{and} \nonumber \\
\beta &=& 2 N \lambda  \,\topt \; \arctan( \wc \topt),
\label{eq:conditions_PC_case}
\end{eqnarray}
where the second one is a transcendental equation. Expanding it to second order in $t$, which is justified by the results of \cite{Smirne2016}, we find $\topt \approx \sqrt{\beta/2 \lambda \wc N}$ which results in the optimal precision as
\begin{eqnarray}
\Delta^2\hat{\omega}\cdot T \;&\gtrsim &\; \sqrt{\frac{2\lambda\wc}{\beta N^{3}}} \, \mathrm{e}^{\sqrt{\frac{2\lambda N}{\beta \wc}} \arctan \left(\sqrt{\frac{\beta \wc}{2\lambda N}}\right)}  \left(1+\frac{\beta \wc}{2\lambda N}\right)^{-\frac{\lambda N}{\beta \wc}} \nonumber \\
&\rightarrow& \sqrt{\frac{2 \lambda \wc \mathrm{e}}{\beta}} \;\frac{1}{N^{3/2}} \;\; (N\rightarrow \infty),
\end{eqnarray}
and is indeed scaling according to the Zeno limit.
Additionally, it is possible to show that an infinitely short Zeno regime immediately yields the SQL. Therefore, remember that the model is described by a Lindblad equation when taking the limit $\wc \rightarrow \infty $, which reduces the correlation time of the environment to zero. Estimating these limits in Eq.~\eqref{eq:transition_prob_PC_case} and \eqref{eq:conditions_PC_case}, we obtain $\topt=\beta/\pi\lambda N$ and the optimal precision is scaling according to the SQL, $\Delta^2 \hat{\omega} \cdot T \geq \pi e \lambda/\beta N$.

\textit{Remark.---} Note that the Zeno scaling can also emerge non-asymptotically, when the function dictating the transversal contraction of the Bloch sphere is always of second order in time.
Then the scaling is immediately Zeno-like, e.g. for Gaussian envelopes, as they are encountered in nitrogen-vacancy centers \cite{Haase2018}.
\begin{figure}[t!]
\includegraphics[width=\columnwidth]{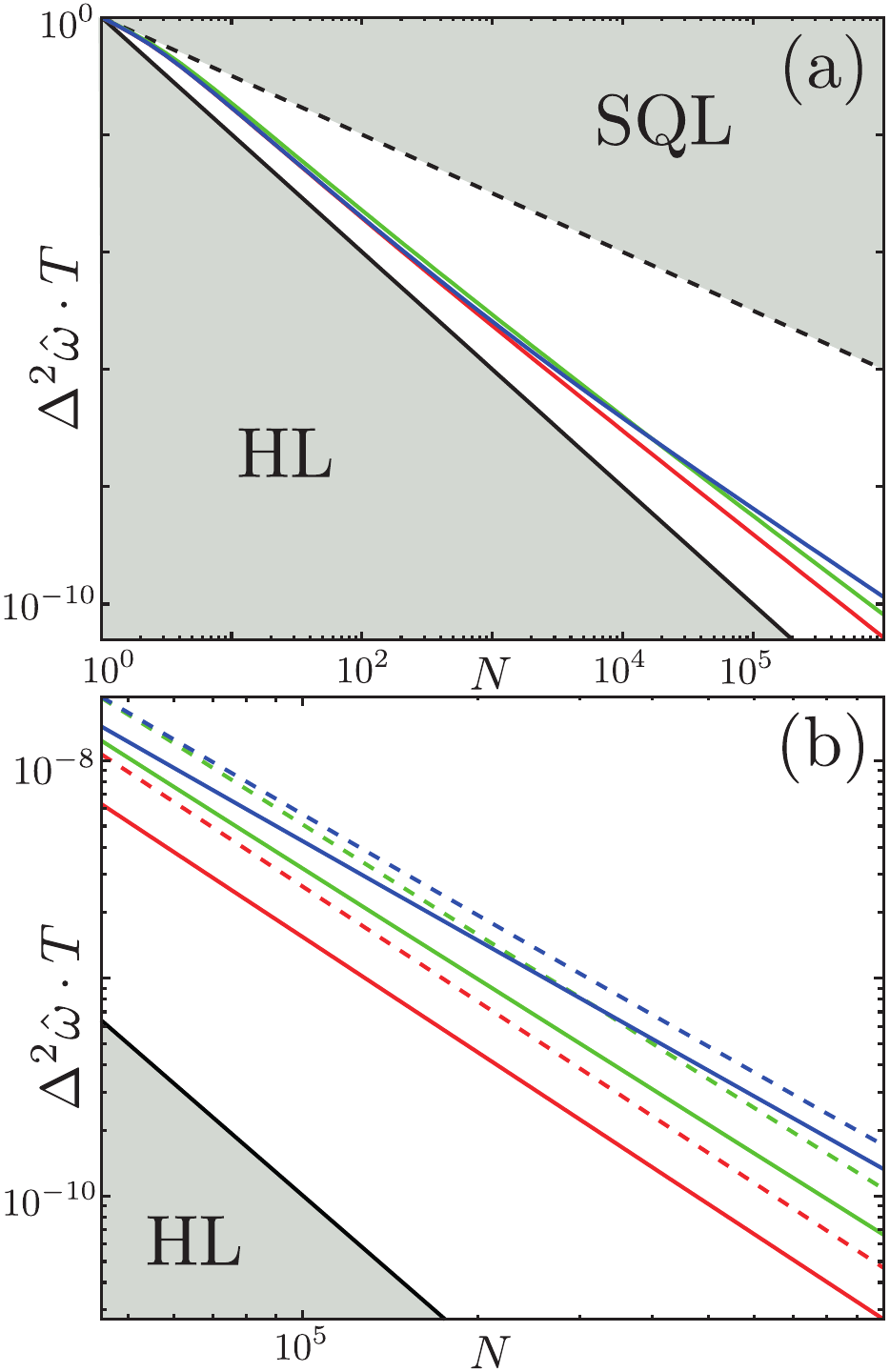}
\caption{\textbf{Scaling of the QCRB in the microscopic noise model.} Panel (a) illustrates the numerically determined scalings of the QCRB in the case of transversal noise [employing Eq.~\eqref{eq:upperBoundcQFI}]. The asymptotic scaling in the semigroup case, given by Eq.~\eqref{eq:scaling_Perp_SG} ($\sim N^{-5/3}$), is shown in green, the non-semigroup scaling given by Eq.~\eqref{eq:scaling_Perp} ($\sim N^{-7/4}$) in red. The solid black line marks the HL ($\sim N^{-2}$), the dashed black line the SQL ($\sim N^{-1}$), hence the white cone represents the region not accessible by classical strategies. The ZL ($\sim N^{-3/2}$) is shown in blue and supports the claim made in Eq.~\eqref{eq:scaling_ZL_NPC}, showing the precision for a noise angle of $\Cang=\pi/100$. Furthermore, it sets the lower bound for PC noise [see Eq.~\eqref{eq:scaling_ZL}]. Note that indeed the limits need to be understood in an asymptotic sense, as the initial increase is slower when additional probes are used. All plots are normalized by their value at $N=1$ such that possible constants are neglected.
Panel (b) represents an excerpt from panel (a) and contains also the scalings of the CRB for a parity measurement according to Eq.~\eqref{eq:parityPrecision}. The curves are plotted with dashed lines while the colors are chosen equivalently to (a). Observe here how the small dephasing component in the blue curves dominates the asymptotic behavior, but for less probes the scaling is closer to the perpendicular cases.}
\label{fig:Scalings}
\end{figure}
\subsection{Transversal noise}
\label{sec:Transversal noise}
A special case is set by noise which is perpendicular to the direction of the frequency encoding (normally chosen as $z$). Indeed, for the model presented this corresponds to $\Cang=0$, but more general one speaks about perpendicular noise at the level of the ME, i.e., whenever the dissipator is of the form
\begin{equation}
\mathcal{D}_t[\rho] = \gamma(t) \left( \alpha_x \sx \rho \sx + \alpha_y \sy \rho \sy - \rho\right),
\end{equation}
with $\alpha_x+\alpha_y=1$, it induces transversal (or perpendicular) noise. Specifically, for a constant rate $\gamma(t)$, this dissipator was analyzed in \cite{Chaves2013} and it was found that the ultimate precision is improved beyond the Zeno limit, yielding
\begin{eqnarray}
\Delta^2\hat{\omega}_{\perp,\mathrm{SG}}\cdot T &\;\gtrsim\;& \frac{1}{N^{5/3}}, \notag \\
\topt^{\perp,\mathrm{SG}} &\sim & N^{-1/3}.
\label{eq:scaling_Perp_SG}
\end{eqnarray}
Crucially, perpendicular noise is not phase covariant, i.e. the condition in Eq.~\eqref{eq:PC_condition} does not hold. In particular, the inclusion of PC breaking terms (which are exactly the ones neglected by the secular approximation \cite{Lostaglio2017, Haase2017}), allows for non-isotropic contractions of the Bloch sphere in $x$ and $y$ direction (see also Fig.\ref{fig:PCspheres}). \textit{Non-phase-covariant} (NPC) dynamics then become sensitive to the initial phase of input states of the cFEP \cite{Haase2017}, which is fixed by the relation of $\alpha_x$ and $\alpha_y$.  Indeed, a dependence of the initial phase was also predicted in \cite{Brask2015}, where the noise model was applied to a specific setup in atomic magnetometry \cite{Wasilewski2010}.

Recently, using the upper bound on the cQFI in Eq.~\eqref{eq:upperBoundcQFI} it was shown numerically, that under the dynamics induced by the ME in Eq.~\eqref{eq:masterEquationohmht} with $\Cang=0$ the precision is ultimatively bounded by \cite{Haase2017}
\begin{eqnarray}
\Delta^2\hat{\omega}_{\perp}\cdot T &\;\gtrsim\;& \frac{1}{N^{7/4}}, \notag \\
\topt^{\perp} &\sim & N^{-1/4}.
\label{eq:scaling_Perp}
\end{eqnarray}
Analogously to the Zeno limit, this scaling emerges under the deviation from pure Lindblad dynamics. Importantly, both scalings, Eq.~\eqref{eq:scaling_Perp} and \eqref{eq:scaling_Perp_SG}, are reached by a parity measurement of GHZ states, see Figs.~\ref{fig:Scalings} (a) and (b). The latter scaling, $N^{-7/4}$ is the best one so far achieved using the cFEP employing the independent noise model.

\subsection{Arbitrary, non-phase-covariant noise}
\label{sec:ZL_NPCnoise}
The dynamics under non-phase-covariant, non-transversal noise are a mostly unexplored category so far when regarded in the context of frequency estimation. It is not too long, that these types of dynamics became important, as the secular approximation performed in the master equation has been a rather standard procedure. Recent technological advances however presented methods to access timescales of the system's dynamics where the contribution of the non-secular terms is not averaged out.

Since an NPC generator does not possess a specific form [conversely they are defined by not being PC, i.e., not of the form in Eq.~\eqref{eq:PC_ME}], it is involved to derive analytic results for the QCRB. Indeed, so far only numerical evidence has been presented, namely bounding the dynamics induced by Eq.~\eqref{eq:masterEquationohmht} in terms of the inequality \eqref{eq:upperBoundcQFI}. It was found that, the ultimate scaling for any $\frac{\pi}{2} \geq \Cang > 0$ may also be given by the Zeno limit iff the decay rate $\gamma(t)$ is time dependent \cite{Haase2017}, i.e.,
\begin{equation}
\Delta^2\hat{\omega}_{\mathrm{Zeno,NPC}} \cdot T  \;\gtrsim\; \frac{1}{N^{3/2}}.
\label{eq:scaling_ZL_NPC}
\end{equation}
We emphasize, that any infinitesimal deviation from $\Cang = 0$ immediately yields the latter scaling, see also Fig.~\ref{fig:Scalings} (b). One observes that in such a case there is always a contribution of the noise in the direction of the parameter imprinting, i.e., a dephasing contribution. Indeed, pure dephasing is \textit{always} PC, hence it should limit the precision as explained in Sec.~\ref{sec:ZL_PCnoise} and derived in \cite{Smirne2016}. Indeed, there it was also shown that the information content in the FI is directly proportional to the length of the Bloch vector's projection into the $xy$ plane, and it was argued that pure dephasing is indeed the most detrimental noise in the estimation scheme. In other words, pure dephasing contributions are the limiting noise factors and when additional noise is added to the (even arbitrary small but not negligible dephasing) dynamics, the precision cannot increase. While this seems intuitive, we want to stress that this must not be the case when the asymptotic limit is not reached or when one has a probe-independent constant improvement in mind. In particular, it was shown \cite{Haase2017} that NPC contributions can increase the single probe QFI on short times when the considered model is kept slightly more general.

\subsection{Motivating toy model}
\label{sec:ToyModel}
\begin{figure}[t!]
\includegraphics[width=\columnwidth]{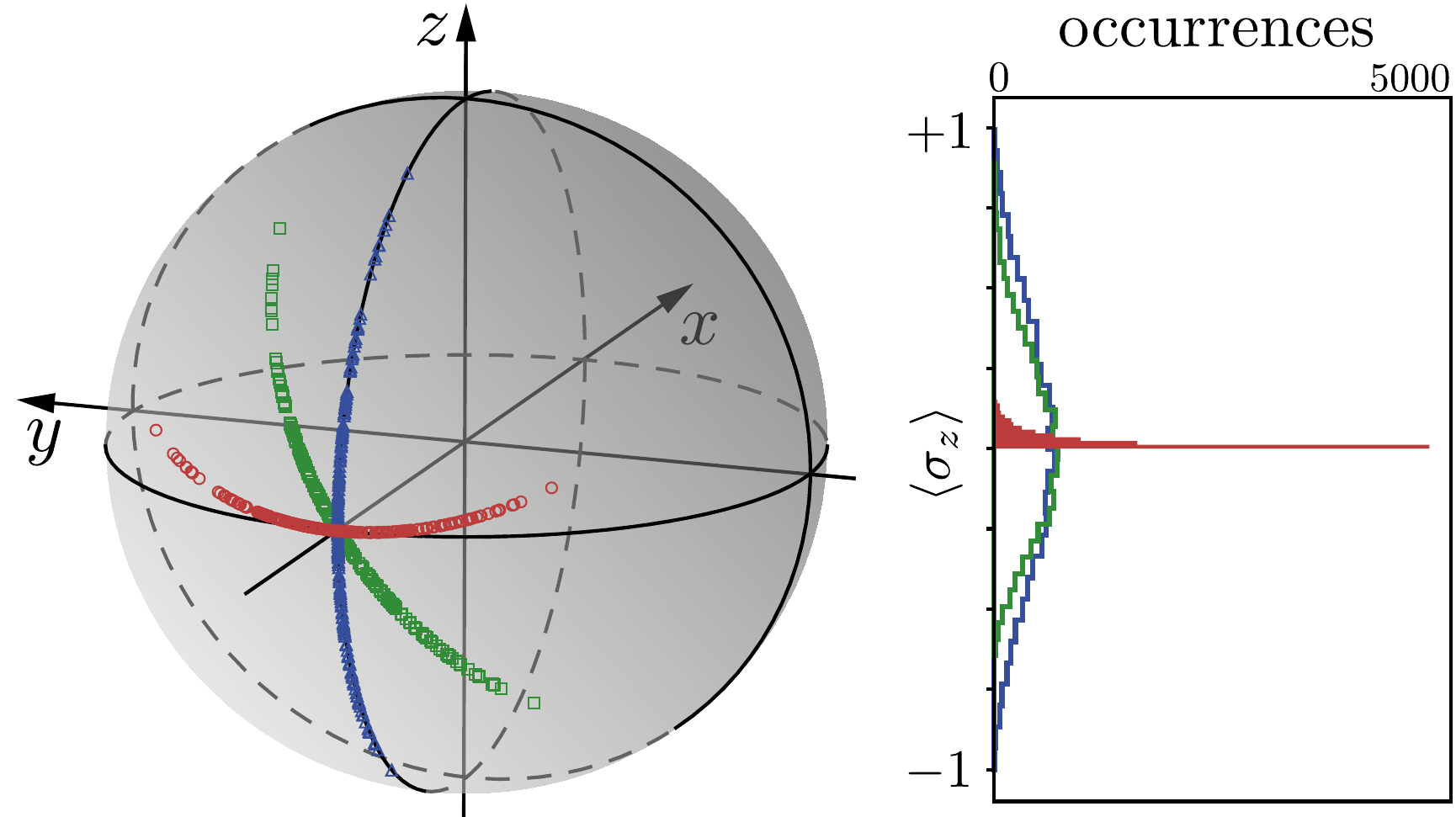}
\caption{\textbf{Toy model for the different directions of the noise.} After 10000 repetitions of the protocol described in the main text (after the second $\pi/2$ rotation), the different outcomes of $\langle \sigma_z \rangle$ are marked with blue triangles ($\theta=\pi/2$), red squares ($\theta=0$) and green circles ($\theta=\pi/4$). Every $x$ component of the noise is along the positive $x$ axis. The right panel shows a histogram of the appearances of the expectation value. Note the much narrower distribution for the transversal case.}
\label{fig:ToyModel}
\end{figure}
To motivate the presented limits pictorially, we dedicate this section to a simple toy model. More precisely, the aim is to illustrate why transversal noise yield an improved scaling and any pure dephasing contribution automatically fixes the precision to be Zeno limited.

Therefore recall that the Ramsey sequence effectively measures the evolved phase of a coherent input state and the precision is fixed by the variance of a corresponding observable which is measured afterwards. For qubit probes, the variance of Pauli operators is completely determined by their expectation value. Let's assume that the dynamics during the free evolution is governed by the Hamiltonian
\begin{equation}
H = \frac{\w}{2}\sz + \eta \left[ \sx \cos (\theta) + \sz \sin(\theta)\right],
\label{eq:toyHam}
\end{equation}
where $\eta$ is the amplitude of a noise process. If an experiment is performed, an amplitude is chosen at random, but it is fixed for the different runs of the experiment (corresponding to $\nu$ in Eq.~\eqref{eq:CRB}). We assume that the drawn amplitudes have zero mean and are distributed according to a Gaussian distribution. We interpret the measured expectation as a single realization drawn from the distribution of possible outcomes.
In the experiment, for each run we prepare an equally weighted superposition of the $\sz$ eigenstates analogously to Sec.\ref{sec:FEP_analyzingDevice}. Let us now consider 3 cases: (i) For pure dephasing we have $\theta=\pi/2$, (ii) for transversal noise we have $\theta=0$ and for a third case (iii) we assume $\theta=\pi/4$. We simulate the evolution via the unitary generated by the Hamiltonian \eqref{eq:toyHam} such that we have a total evolution of $\w t = \pi/2$. Every result plotted on the Bloch sphere shown in Fig.\ref{fig:ToyModel} corresponds to the outcome of a different experiment. A $\pi/2$ rotation around the $x$ axis transforms the phase information of the states onto a population difference, where we can extract the distribution of $\langle \sigma_z \rangle$, see Fig.~\ref{fig:ToyModel}, which show a drastically reduced variance for case (ii) when compared to the cases (i) and (iii), which themselves yield pretty similar results. Note that for case (i), the non-negative expectation value is a result of the particular geometry in combination with small values of $\eta$.

To that end, this gives a motivation for why transversal noise is a special case, while any longitudinal component will reduce the scaling to the ZL.

%
%%%%%%%%%%%%%%%%%%%%%%%%%%%%%%%%%%%%%%%%%%%%%%%%%%%%%%%%%%%%%%%%%%%%%%%%%%%%%%%%%%%%%%%%%%%%%%%%%%%%
\subsection{Remarks}
\subsubsection{The role of non-Markovianity}
\begin{figure}[t!]
\includegraphics[width=\columnwidth]{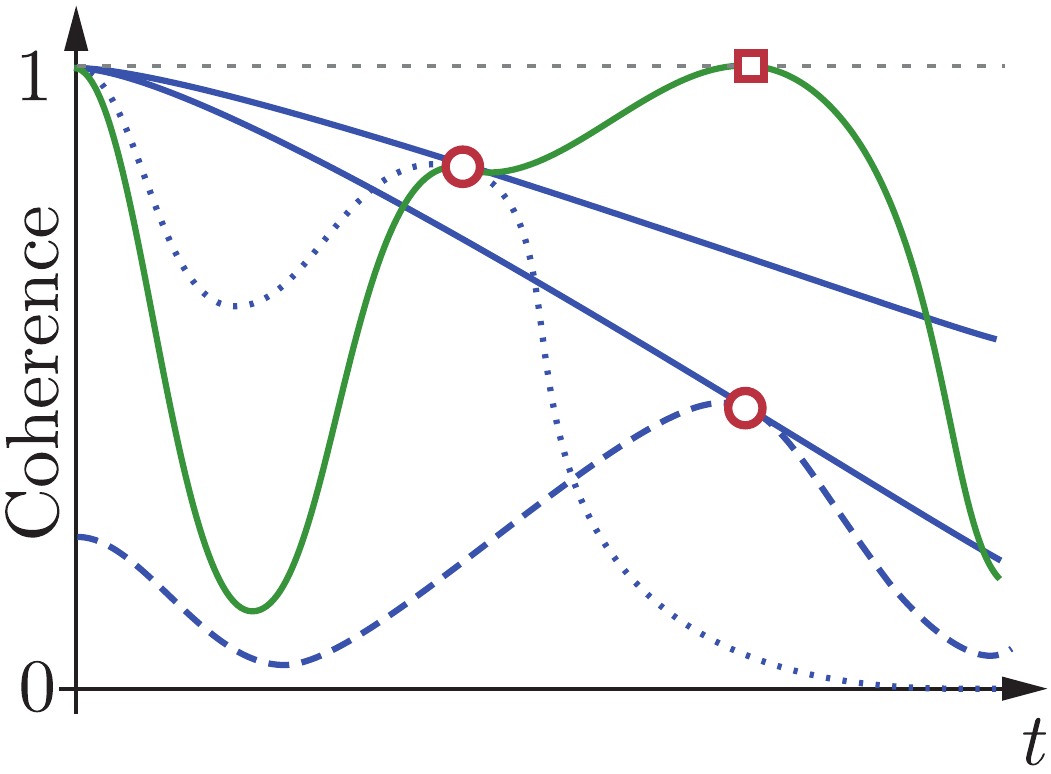}
\caption{\textbf{Role of non-Markovianity in pure dephasing.} Each optimal point (red circles) of a non-Markovian decay process (dashed and dotted blue lines) can be reached by Markovian process (blue solid lines), therefore demonstrating that the non-Markovianity of the dynamics cannot provide an advantage over Markovian dynamics. The only exception is the case of a full revival as shown by the green curve and marked by the red square. This point cannot be reached via a Markovian dynamics and decay rates $\gamma_r>0$.}
\label{fig:NM}
\end{figure}
The role of non-Markovianity in topics referring to quantum metrology is, by far, not sorted out yet. However, we stress that for the configuration of the cFEP non-Markovianity does not play any role when it comes to the ultimate limits in the asymptotic regime. A detailed analysis and the proof for phase covariant noise is given in \cite{Smirne2016}, but we give an intuition in the following.

As shown in \cite{Smirne2016} and argued in Sec.~\ref{sec:ZL_PCnoise}, performing measurements at shorter and shorter time scales is crucial in order to overcome the SQL. This not only implies that, as said, the key property is the violation of the semigroup composition law on short time scales (rather than a specific form of non-Markovianity), but one can also show that, apart from the unrealistic case of a full revival, performing a measurement on longer time scales (e.g., waiting for a back-flow of information) would be in any case detrimental and furthermore reducing the scaling of the error to the SQL. Now, such a strong result is certainly a consequence of the asymptotic regime $N\rightarrow \infty$ taken into account in \cite{Smirne2016}. However, one can easily argue that non-Markovianity is not really a necessary resource for the FEP, even in the finite-$N$ regime.

Crucially, note that all the FIs defined above are local quantities in time, i.e., they can not capture any temporal correlation in the evolution of the state. In other words, they only take the instantaneous state of the system into account. E.g., as mentioned already in Sec.~\ref{sec:ZL_NPCnoise}, the achievable precision depends on the available coherence orthogonal to the imprinting of $\w$. Any dynamics, whose value of coherence coincide at a given point yield the same cQFI, which means that even if the cQFI increases in time during one evolution due to non-Markovianity, one will always find a different Markovian dynamics reaching the same QFI at the same time and thus providing the same precision. This argument is illustrated in Fig.~\ref{fig:NM}.
Nevertheless, we stress that non-Markovianity can be of course of practical advantage, given a specific setting and hence a restricted set of available dynamics.

Moreover, note that the temporal derivative of the QFI has been proposed as a measure of non-Markovianity as it quantifies the information flow between the system and the environment \cite{Lu2010}.
\subsubsection{Precision, accuracy and sensitivity}
\begin{figure}[t!]
\includegraphics[width=\columnwidth]{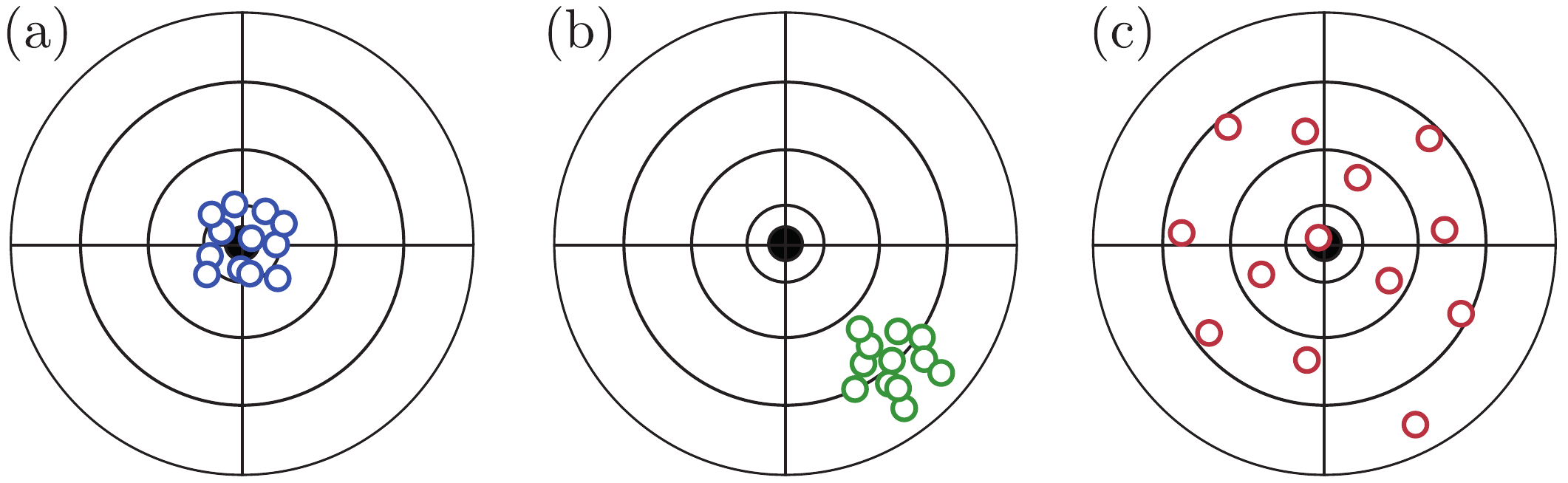}
\caption{\textbf{Precision and accuracy.} Imagine throws onto a dartboard, each consecutive hit is marked with a circle. The player of board (a) is accurate and precise, as his throws have a small spread and are distributed around the center. The player on (b) is very precise but lacks accuracy. His throws also possess a narrow distribution, however around a point which is displaced from the center of the target. The throws of the last player on (c) are evenly but widely distributed around the center, therefore he is accurate without possessing any precision. To connect this illustration with the estimation task treated in this work, every throw onto the dartboard has to be associated with a whole experiment which is conducted.}
\label{fig:precision}
\end{figure}
In this paragraph we would like to clarify some terms commonly encountered in the literature and often used interchangeably.
The notions of an accurate and precise measurement can be linked to properties of the estimator \cite{Bos2007}. Any unbiased estimator is accurate. Therefore, asymptotic unbiasedness also guarantees an accurate measurement in the asymptotic regime. The notion of precision is surely connected to the variance of the estimator and for unbiased estimators it is equal to the MSE. Importantly, precision is a term describing the closeness of results obtained by repeated performances of the experiment (as long as the true value does not change), the results may still be biased away from the true value, see the illustration in Fig.~\ref{fig:precision}. Note that, any efficient estimator is accurate and precise as it is consistent. In particular, any MLE is precise and accurate in the asymptotic regime.

A further term often used when quantifying the performance of quantum sensing experiments is \textit{sensitivity} \cite{Taylor2008, Degen2017}. Note that this term may be misleading in a broader context, since other communities use the term \textit{noise equivalent power} (NEP) $\eta$ \cite{Richards1994, Boreman1998}, while sensitivity is then referred to the slope of the response curve \cite{DAmico2001, Fraden2004}.
NEP is a measure of the \textit{signal-to-noise ratio} (SNR), specifically it is defined as the signal yielding the $\mathrm{SNR}=1$. Since this depends on the resources at hand, one usually chooses a fixed total time of $T=1\,\mathrm{s}$.
For a response curve or signal $S_T(\w)$ and the noise $\sigma_T(\w)$ we have the SNR
\begin{equation}
\mathrm{SNR} = \frac{S_T(\w)}{\sigma_T(\w)}.
\end{equation}
Here we include the possibility to repeat measurements and denote the affiliation to the specific total time in the index $T$. This may change the response itself but crucially the repetitions reduce the noise, typically by a factor $\sqrt{T/t}$ where $t$ is the duration of a single run, compare the discussion in Sec.~\ref{sec:motivation}. We emphasize that in practical applications $\sigma_T(\omega_0)$ is a sum of noise contributions from different sources, e.g. electrical noise, counting errors of quantized signals like photons or precisely the quantum shot noise (or projection noise) \cite{Itano1993}.

In principal the NEP is given by $S_T(\w)=\sigma_T(\w)$ for $T=1s$, however it is convenient to express it in terms of the quantity to estimate. Since the sensor needs to be calibrated to a known reference point $\omega_L$, we express the response as (recall $\w = \omega_L +\delta\omega $)
\begin{equation}
S_T(\w) \approx S_T(\omega_L) + \left. \frac{\partial S_T(\w)}{\partial \w} \right \vert_{\w=\omega_L} \delta\omega,
\label{eq:response_locality}
\end{equation}
and therefore we arrive at the NEP (note that the $\sqrt{\mathrm{Hz}}$ is required to obtain the desired units of $\eta$)
\begin{equation}
\delta\omega = \sigma_T(\w) \left[\left| \left.\frac{\partial S_T(\w)}{\partial \w}\right\vert_{\w = \omega_L}\right|\right]^{-1}\Bigg\vert_{T=1\,\mathrm{s}} = \eta\,\sqrt{\mathrm{Hz}}.
\label{eq:sensitivity}
\end{equation}
Thereby the value of $S_T(\omega_L)$ is a bias which is known via the calibration and we set it to zero without any loss of generality and we prefer $\delta \omega$ to be positive, hence we take the absolute value of the derivative. The NEP shares similarities with the error propagation Eq.~\eqref{eq:errorPropagation} and we may indeed interpret $\delta \omega$ as the upper limit in precision when we understand the sensing experiment as a task of frequency matching, where we aim to tune $\omega_L$ as close as possible to $\w$. Particularly, $S_T(\w)$ is then the expectation value of a quantum mechanical observable and $\sigma_T(\w)$ its standard deviation. Furthermore, the requirement of local estimation is implied by Eq.~\eqref{eq:response_locality} where the derivative is the sensitivity of the sensor. However, as mentioned, be aware that some communities refer to the concept of NEP as sensitivity of the sensor and with that terminology, \textit{responsivity} is used for the local slope \cite{Boreman1998}.

For illustration purposes, let us again derive the NEP (i.e. the sensitivity) of a single probe Ramsey experiment as used for magnetometry of a magnetic field, i.e., $\w = \gamma_\mathrm{a} B$ is determined by the Zeeman interaction of the atomic probe (gyromagnetic ratio $\gamma_\mathrm{a}$) with a magnetic field with amplitude $B$. Here, $S_T(\w)$ is given by the survival probability of the initial state, Eq.~\eqref{eq:pLindbladSep} and $\sigma_T(\w)$ represents the shot noise. As already mentioned, the shot noise is given by $\sigma_T(\w)=\sqrt{S_T(\w)[1-S_T(\w)]/\nu}$ with $\nu = T/t$. Plugging these expressions into Eq.~\eqref{eq:sensitivity} yields
\begin{eqnarray}
\delta\omega = \left. \frac{\sqrt{e^{-2\gamma t}-\cos^2(\omega t)}}{\sqrt{t T}\abs{\sin(\omega t)}}\right \vert_{T=1\,\mathrm{s}},
\end{eqnarray}
which is minimized for $t=\pi/(2\omega)$. Further minimizing over $t$ yields the optimal time $t_{\mathrm{opt}}=1/(2\gamma)$ and translating this into the NEP for the field amplitude yields
\begin{eqnarray}
\delta B &=& \sqrt{\frac{2 e}{\gamma_\mathrm{a}/\gamma}} \frac{1}{1\,\mathrm{s}} = \sqrt{\frac{2 e}{\gamma_\mathrm{a} T_2^*}} \sqrt{\mathrm{Hz}} \notag \\
\Rightarrow \eta &=& \sqrt{\frac{2 e}{\gamma_\mathrm{a} T_2^*}}\; \mathrm{and} \; [\eta] = \mathrm{[B]} \sqrt{\mathrm{Hz}}^{-1}
\end{eqnarray}
In the last step we used that the coherence time of the probe is given by $T_2^*=1/\gamma$. Note that the NEP $\eta$ is given in terms of the units of the parameter (here $B$) divided by $\sqrt{\mathrm{Hz}}$ giving a reference to the integration time of $T=1\,\mathrm{s}$.
\subsubsection{Ultimate precision without entanglement}
\label{sec:SubSQL_HO}
As we have seen, in parallel estimation strategies a necessary condition to overcome the SQL with respect to the number $N$ of probes is the entanglement among the latter. However, it has been shown that the same precision can be achieved in a sequential strategy where, instead of $N$ initially entangled probes, one has an $N$-step protocol with one single probe \cite{Luis2002, Higgins2007}. For the case of a unitary operator $\exp\left(-it\w\sz/2\right)$ which is applied $N$ times to an initially equally weighted superposition of $\ket{0}$ and $\ket{1}$, one obtains the state of the probe after the interrogation time as $(\ket{0}+e^{-iN\w t}\ket{1})/\sqrt{2}$ and the survival probability is hence given by
\begin{equation}
p_{\w,t} = \frac{1+\cos(N\w t)}{2},
\end{equation}
yielding the scaling $1/N^2$ of the precision. However, note that in practice such a protocol is also challenging to implement, as the setup has to stabilized, also against noise, during the total duration $Nt$ of the experiment. 

On the other hand, in case of a bosonic systems whose indistinguishable particles constitute the probe, it is more natural to treat them all together as an isolated quantum system. Then, from such a perspective, one may interpret the HL to be attainable with $N$ bosonic particles, i.e., $N$ excitations of a single bosonic mode (given a perfect phase reference) \cite{Benatti2013, Braun2017}. However, one has to note that such states still carry particle entanglement which is, contrary to mode entanglement, necessary to obtain a quantum advantage in non-sequential schemes. Nevertheless, these details goes far beyond the scope of this tutorial but can be found in \cite{Killoran2014, DemkowiczDobrzanski2015}.
\subsubsection{Geometrical distance of quantum states}
Let us briefly note the connection between the (quantum) Fisher-Information and the distinguishability of different quantum states. Therefore note that, besides in this work we are focusing on frequency estimation, the achievable precision for any other parameter $\lambda$ can be analyzed using the formalism presented here by making the identification $\w \mapsto \lambda$ in the Eqs.~\eqref{eq:CRB} and \eqref{eq:QCRB} \cite{Note5}.

As we already noted during the introduction, the problem of a finite estimation precision is emerging from the fact that probability theory is involved in the performance of measurements.
Based on classical probability theory, in \cite{Wootters1981} a notion of statistical distance between two probability distributions was introduced. If one parameterizes these distributions as $p_{\lambda}$, the distance between $p_{\lambda_1}$ and $p_{\lambda_2}$ can be defined as the shortest path between the two, calculated in the space of all $p_{\lambda}$. An intuitive measure of the length is given in terms of the probabilities which can be distinguished along the path. In the case that $p_\lambda$ is referred to $N$ possible outcomes, the length (note the appearance of the FI)
\begin{equation}
l = \frac{1}{2}\int_{\lambda_1}^{\lambda_1} \mathrm{d}\lambda \left \lbrace \sum_{n=1}^N \frac{1}{p_{\lambda}(n)} \left[ \frac{\mathrm{d}p_{\lambda}(n)}{\mathrm{d}\lambda}\right] ^2 \right \rbrace ^{\frac{1}{2}}
\end{equation}
is minimized to yield the statistical distance
\begin{equation}
d(p_{\lambda_1},p_{\lambda_2}) = \arccos \left( \sum_{n=1}^{N} \sqrt{p_{\lambda_1}(n)\,p_{\lambda_2}(n)} \right).
\end{equation}
This result has then been transferred to the quantum regime where an $N$ dimensional pure state $\ket{\psi_\lambda}$ is measured. Crucially, here the probabilities $p_{\lambda}(n)$ will depend on the chosen measurement basis and hence a further optimization can be performed. In particular, the optimal measurement basis includes one of the states itself, which yields
\begin{equation}
d\left(\ket{\psi_{\lambda_1}},\ket{\psi_{\lambda_2}}\right) = \arccos \abs{\braket{\psi_{\lambda_1}}{\psi_{\lambda_2}}}.
\end{equation}
Therefore, the distinguishability directly relates to the angle enclosed by two states in the Hilbert space. This result can be transformed into a metric for neighboring pure states $\ket{\psi_\lambda}$ and $\ket{\psi_{\lambda+\delta\lambda}}$ giving the Fubini-Study metric \cite{Gibbons1992} and in \cite{Braunstein1994} a generalization for mixed states is presented. The statistical distance for two states is then given by the Bures distance \cite{Bengtsson2006}
\begin{equation}
d_B(\rho_{\lambda_1},\rho_{\lambda_2}) = \arccos F(\rho_{\lambda_1},\rho_{\lambda_2}) = \arccos \tr{\sqrt{\sqrt{\rho_{\lambda_1}}\rho_{\lambda_2}\sqrt{\rho_{\lambda_2}}}}
\end{equation}
where $F$ denotes the Fidelity \cite{Nielsen2010}. Interestingly, for neighboring states this equation can be expanded yielding
\begin{equation}
d_B(\rho_{\lambda_1},\rho_{\lambda_2}) = \frac{1}{2}\sqrt{F_Q[\rho_\lambda]} \, \delta \lambda+ O(\delta \lambda ^2),
\end{equation}
which shows the connection between the QFI and the notion of statistical distance between quantum states.
\section{Outlook beyond the independent noise model: Correlations and control}
\label{sec:beyond_indep_noise}
By now it should be clear that the cFEP employing the independent noise model is an idealization. In addition, the bounds mentioned in this tutorial may be the ultimate bounds in an asymptotic regime, however, current realizations of the protocol in experimental setups struggles to achieve this regime. On top, there can be initial correlations in the noise affecting the individual probes and the probes may also interact with each other in principle. Furthermore one can think of control methods during the interrogation time, which may suppress noise, perform error correction or increase the sensitivity to the frequency to be estimated. The details of these techniques go beyond the scope of this work, however we want to complete it by mentioning the recent progress in the field.
\subsection{Correlated noise and interacting probes}
The cFEP sets fixed requirements onto the setup to be analyzed. Indeed, the boundary conditions of independent and identical noise are rather an idealization. Despite the fact that this provides an accurate description of the noise in many circumstances, there are certainly situations of interest where correlations of the noise are actually relevant. On top, the different probes are prohibited to interact during the interrogation time, also a necessity which is not always given. Especially in the context of probes which are desired to be prepared in an entangled state, where corresponding methods relying on the inter-probe interaction exist.

However, it is not a priori given that these flaws in a realization of the cFEP are a disadvantage. Considering pure dephasing, it was shown that noise which is spacially correlated along the used probes can beat the SQL with Lindbladian \cite{Dorner2012, Jeske2014, Guo2016} and non-Lindbladian noise \cite{Yousefjani2017}. In particular, it was shown that for some antisymmetric entangled preparations of the input state, the correlations in the noise allow for the identification of decoherence free subspaces (DFS) which in turn even allow for the restoration of the HL. In \cite{Jeske2014} it was calculated that the spacial length over which the correlations decay is crucial for the achievable scaling and the HL manifests for correlation lengths longer than the chain of probes. This infinite correlation length was implicitly assumed in \cite{Dorner2012} where a linear chain of trapped ions was investigated. To exemplify the latter, consider a master equation describing the total even $N$-probe state under correlated dephasing which may take the form
\begin{eqnarray}
\frac{\mathrm{d}}{\mathrm{d}t} \R &=& - i \left[ H,\R \right] + \gamma \left( V \R V - \frac{1}{2} \left \lbrace V^2,\R \right \rbrace \right), \;  \\ &&\mathrm{when, e.g.,} \notag \\
H &=& \frac{\omega_1}{2}\sum_{n=1}^{N/2} \sz^{(n)} + \frac{\omega_2}{2}\sum_{n=N/2+1}^{N} \sz^{(n)} + \xi(t) \sqrt{\gamma} \sum_{n=1}^N \sz^{(n)} \notag
\end{eqnarray}
where $V = \sum_{n=1}^N \sz^{(n)}$ and $\xi(t)$ is a delta correlated, zero mean stochastic process, i.e., white noise. Note, that a state which is part of the DFS has to satisfy $V\ket{\psi_{\mathrm{DFS}}(t)}=0$ at all times. One way to construct such a subspace is the following. Therefore, note that the first two terms in the Hamiltonian can be rearranged as
\begin{equation}
H_0 = \frac{\omega_1-\omega_2}{4} \left( \sum_{n=1}^{N/2} \sz^{(n)} - \sum_{n=N/2+1}^N \sz^{(n)}\right) + \frac{\omega_1+\omega_2}{4} V.
\end{equation}
Then, an input state of the form \cite{Dorner2012}
\begin{equation}
\ket{\psi} = \frac{1}{\sqrt{2}}\left( \bigotimes_{n=1}^{N/2}  \ket{1}  \bigotimes_{n=N/2+1}^{N} \ket{0} +\bigotimes_{n=1}^{N/2}  \ket{0}  \bigotimes_{n=N/2+1}^{N} \ket{1} \right)
\end{equation}
fulfills the DFS criteria and can be used to measure the frequency $\w=\omega_1-\omega_2$.
Interestingly, if one does not exploit the existence of theses DFS, under the conditions of correlated noise GHZ states dephase on a timescale $\propto N^{-2}$ compared to an uncorrelated preparation when employed on conventional Ramsey spectroscopy, i.e., all ions possess an equal splitting. This effect was called \textit{superdecoherence} \cite{Monz2011}, implying that GHZ states are strongly disadvantageous. Indeed, it was found that the precision using GHZ states is then independent of $N$, and furthermore, for optimized input states it was demonstrated that a constant, $N$ independent part prevails in the precision, i.e.,
\begin{equation}
\Delta^2 \hat{\omega} \approx \frac{\gamma C_1}{T} + \frac{\gamma C_2}{T N^{1.8}},
\end{equation}
where $C_1$ and $C_2$ are some constants determined numerically \cite{Dorner2012}. To that end, the example presented may suggest that the assumption of local noise in the cFEP is an optimistic one, yielding a better precision than for correlated noise. On the other hand, this is no longer true in the special case of an appearing DFS where finally the HL can be reached. In any case, the precise comparison between the two scenarios is under investigation.

Another, until now only briefly investigated scenario are probes interacting among each other.
Whether the parameter independent interaction of the probes can increase the precision is yet to be fully explored. It was shown that the estimation of a transverse field in an Ising-Hamiltonian can be performed with Heisenberg limited precision \cite{Skotiniotis2015} and similar results have been derived for estimation procedures close to phase transitions \cite{Rams2018}. Furthermore, there are investigations for the case when the frequency to be estimated is given by the coupling constant of $k$-body interactions. Precisely, the total encoding Hamiltonian has the form
\begin{equation}
H = \w \left(\sum_{n=1}^{N} h_0^{(n)} \right)^k
\end{equation}
where $h_0^{(n)}$ is the same operator for each probe. Such a case is clearly operating outside the framework of the cFEP described until here, as the best precision achievable under such evolution scales as $N^{-2k}$ \cite{Boixo2007, Demkowicz2017, Beau2017}. Remarkably, for initial product states this scaling is only slightly altered to $N^{-(2k-1)}$ and in specific cases it is enough to consider separable measurements to achieve the optimal scaling, while the scaling is also maintained under Lindbladian dephasing \cite{Boixo2008}. An experiment involving Bose-Einstein-Condensates was proposed \cite{Boixo2008Jul} and performed \cite{Napolitano2011}. Despite the simplified preparation of the initial input product state, the experimental difficulty is shifted to the generation of a $k$-body Hamiltonian (in this case $k=2$ was realized). It is worth stressing, that such a scheme also uses exclusive quantum resources as entanglement is generated during the interrogation time. This is in contrast to the cFEP introduced here, where the entanglement is injected during the input state preparation and interaction during the interrogation time is not considered.
\subsection{External control}
A natural approach to an increase of precision is the suppression of noise acting on the probes \cite{Macchiavello2000, Preskill2000}. Within the cFEP, this corresponds to multiple applications of the channel during the interrogation time, but between the channels it is allowed to perform unitary operations. Assuming time-homogeneous Lindbladian noise, bounds under infinitely fast control have been found. In particular for qubit probes, it was shown that rank one Pauli noise can be eliminated completely, as long as it is not parallel to the imprinting of the parameter \cite{Sekatski2017}. Therefore, assume that a probe which evolves according to the semigroup limit ($\omega_C \rightarrow \infty $) of Eq.~\eqref{eq:spinBosonME}, while it is initially in an entangled GHZ state with a noiseless ancilla. We define a logical qubit (the so called ``code space") via the the subspace $\lbrace \ket{\uparrow 1}, \ket{\downarrow 0}\rbrace$ with $\bar{\sigma}\ket{\uparrow(\downarrow)}=\ket{\downarrow(\uparrow)}$ and some arbitrary reference basis $\lbrace \ket{0},\ket{1}\rbrace$ for the ancilla. Integrating the master equation over a small time step $\mathrm{d}t$ and projecting into the code space via $P_c = \ketbra{\uparrow 1}{\uparrow 1} + \ketbra{\downarrow 0}{\downarrow 0}$ and the error space ($P_e = \id - P_c$) yields
\begin{eqnarray}
\rho_C(\mathrm{d}t) &=& \rho(\id - \gamma\, \mathrm{d}t) - i \frac{\w}{2} \left[ \kappa ,\rho\right] \mathrm{d}t + O(\mathrm{d}t^2), \nonumber \\
\rho_E(\mathrm{d}t) &=& \gamma \, \bar{\sigma} \rho \bar{\sigma} \, \mathrm{d}t + O(\mathrm{d}t^2),
\end{eqnarray}
respectively. Note that we implicitly assume the tensor product $\otimes \id_A$ for the ancilla space. In case an error emerges, $\bar{\sigma}$ is applied to $\rho_E(\mathrm{d}t)$, otherwise the system remains unmodified. After the correction, the system is in a mixed state $\rho(\mathrm{d}t) = \rho_C(\mathrm{d}t) + \bar{\sigma} \rho_E(\mathrm{d}t) \bar{\sigma}$. Rearranging the terms, yields the differential equation
\begin{equation}
\frac{\mathrm{d}}{\mathrm{d}t} \rho = -i \frac{\w}{2} \left[ \kappa ,\rho\right]
\end{equation}
in the limit $\mathrm{d}t\rightarrow 0$. This corresponds to a unitary evolution with the effective Hamiltonian $\kappa = \cos^2 \Cang  \, \sz - \sin 2\Cang \,\sx /2$, hence the error correction rotates the encoding basis. However, now the analysis of Sec.~\ref{sec:HL} applies. The eigenvalues of $\kappa$ are given by $\pm \cos \Cang$, representing the only penalty of the scheme, which is a slower encoding of the parameter. Importantly, as soon as $\Cang=\pi/2$, the noise and the encoding are parallel and the error correction also removes all information of $\w$ encoded during $\mathrm{d}t$.

Importantly, this result has been generalized recently to any finite-dimensional probe \cite{Demkowicz2017, Zhou2018}, showing that one can always restore the HL, if the encoding Hamiltonian is not contained in the linear span of the identity $\id$ and the Lindblad operators $V_k,\,V_k^\dagger ,\,V_k^\dagger V_j,\,\forall k,j$. In particular, if there is a dephasing term $H=\alpha V_{\mathrm{dephasing}}$ with some constant $\alpha$ the HL can not be reached, confirming the detrimental role of pure dephasing (see Sec.~\ref{sec:ZL_NPCnoise} and the related discussion).

Relaxing the requirement of a time-homogeneous noise process, it was shown that dynamcial decoupling restores the ZL \cite{Sekatski2016}. However, since dynamical decoupling is limited by the correlation time of the environment \cite{Viola1999}, one resorts to error correction schemes, which in general can be applied on the time scales of the effects of the noise \cite{Shor1995, Nielsen2010, Herrera-Marti2015}. While these base on the idea to prolong the coherence time \cite{Arrad2014, Kessler2014} (with experimental implementation \cite{Unden2016}), limiting processes as spontaneous emission may be corrected by observation of the environment \cite{Plenio2016, Gefen2016}. Furthermore, a way to utilize open quantum systems is the engineering of noise processes which drive the probes back into the code space \cite{Oreshkov2007, Reiter2017}. A different approach was aimed to preserve the QFI itself, rather than the input state \cite{Lu2015}.
Recently, it was also observed that for time-homogeneous processes a continuous measurement \cite{Gammelmark2014, Genoni2017} of the environment \cite{Albarelli2018} can restore the $1/N^2$ scaling of the error. Interestingly, while in a parallel-noise scenario the full noiseless HL can be reached, if the noise is transversal, the noiseless optimal error can be obtained up to a constant factor, showing that in this case the estimation precision is unavoidably lowered by the interaction with the environment, even if all the degrees of freedom of the latter can be accessed. 
On the other hand, a very recent proposal suggests the implementation of fault tolerant strategies, which could provide a different avenue for counteracting the effects of noise  \cite{Kapourniotis2018}.
%
%
%%%%%%%%%%%%%%%%%%%%%%%%%%%%%%%%%%%%%%%%%%%%%%%%%%%%%%%%%%%%%%%%%%%%%%%%%%%%%%%%%%%%%%%%%%%%%%%%%%%%
\subsection{Time dependent encoding}
Recently, interesting progress has been made for the case of time dependent encoding Hamiltonians. Before examining the setting we should stress that in that context the term ``frequency estimation" is often referred to the frequency of an ac-signal \cite{Degen2017} and thus differs from the definition we adapted in this work. Furthermore, instead of estimating the precision for the best scaling in $N$ one is rather interested in the scaling with the available time $T$ which for time independent encodings is usually given as $\Delta^2\omega \sim T^{-2}$, compare Sec.~\ref{sec:motivation},\ref{sec:LindbladDesphasing} and \ref{sec:HL}. However, with time dependent encodings this scaling can be overcome. A trivial example is the Hamiltonian
\begin{eqnarray}
H_f(t)=f(\w,t) \, G
\label{eq:time_dep_ramping_H}
\end{eqnarray}
where $G$ is a time independent hermitian operator and $f(\w,t)$ a real valued function. Employing Eq.~\eqref{eq:QFIpureState}, the QFI yields
\begin{equation}
F_Q[U\ket{\psi}] = 4 \left( \frac{\partial \int_0^t f(\w,\tau) \,\mathrm{d}\tau}{\partial \w}\right)^2\, \Delta^2G\Big\vert_{\ket{\psi}}.
\end{equation}
Obviously, depending on the form of $f(\w,t)$ the precision $\Delta^2 \omega \geq F_Q^{-1}$ can take a different scalings in $t$ or equivalently in $T$. As exploited in \cite{Gefen2017}, the application of a suitable control Hamiltonian to some time dependent encoding may transform the Hamiltonian to the one in Eq.~\eqref{eq:time_dep_ramping_H}.
An elegant way to construct a suitable control Hamiltonian for any time dependent encoding was presented in \cite{Pang2017}, starting from the observation that $F_Q[U\ket{\psi}] = 4 \mathrm{var}[\mathfrak{H}(\w,t)]$, where $\mathfrak{H}(\w,t)=-iU(\w,t)\,\partial U(\w,t)/\partial \w$ and $U(\w,t)$ is the unitary operator generating the evolution governed by an arbitrary time dependent Hamiltonian.
It was shown that the applied control Hamiltonian should be constructed such that it steers the input state on the path on optimal sensing states as the system evolves. Analogously to the time independent case examined in \cite{Giovannetti2006}, this state is always given by an equally weighted superposition of the instantaneous eigenstates of $\mathfrak{H}(\w,t)$ with the instantaneous maximum  ($\mu_{\mathrm{max}}(t)$) and minimum  ($\mu_{\mathrm{min}}(t)$) eigenvalues. Hence at each time $t$ we have $\ket{\psi_t}\propto \ket{\mu_{\mathrm{max}}(t)}+\ket{\mu_{\mathrm{min}}(t)}$ with $\mathfrak{H}(\w,t)\ket{\mu_\mathrm{j}t)}=\mu_\mathrm{j}(t)\ket{\mu_\mathrm{j}(t)}$ and the QFI yields \cite{Pang2017}
\begin{equation}
F_Q\left[\ket{\psi_t}\right] = \left[ \int_0^{t} \mu_{\mathrm{max}}(\tau) - \mu_{\mathrm{min}}(\tau)\, \mathrm{d}\tau\right]^2.
\end{equation}
To consider an explicit example, it was shown that the precision in estimating $\w$ when encoded by $H(t)=-B\left[\sx \cos(\w t)+ \sz \sin(\w t)\right]$ scales as $\Delta^2\omega \geq 1/B^2T^4$, which is then also the best precision achievable for that setting. Contrary, it is worth mentioning that estimating the amplitude $B$, i.e., the frequency we were focusing on in all other chapters of this tutorial, scales as $\Delta^2 B\geq 1/4T^2$. The analytic form of the control Hamiltonian can be found in \cite{Pang2017}, but one should mention that in general this control depends on the frequency to be estimated. While this seems contradictive, it is enough to recall that the estimation is performed locally. Hence, using a close estimate for the frequency in the control Hamiltonian also improves the precision, as can be seen in \cite{Gefen2017}. Furthermore, \cite{Pang2017} showed that one can use an adaptive scheme, where estimations of the parameter are used as a feedback for the control and the quartic scaling is then reached in an asymptotic regime of repetitions. A further example of the application of a control Hamiltonian can be found in \cite{Yang2017}, where the estimation of the speed of a Landau-Zener sweep also shows the quartic scaling in the total time.

A time dependent encoding of the form $H(t)=A\sin(\w t)\sz$ has been studied experimentally, exploiting a nitrogen-vacancy center in diamond \cite{Schmitt2017, Boss2017} or involving a superconducting transmon circuit coupled to a waveguide cavity \cite{Naghiloo2017}. The latter used a control Hamiltonian constructed via the methods in \cite{Pang2017}, indeed showing a scaling $\sim T^{-4}$ for times shorter than the coherence time of the probe. Regarding the scheme employing nitrogen-vacancy centers, a different control was employed where the limiting factor was set by the coherence time of the signal itself. Dividing the total time into small blocks where a dynamical decoupling sequence and a subsequent measurement was performed, the total FI is the sum of the FI of the different measurements, which resulted in a scaling of
\begin{equation}
\Delta^2 \omega \sim \frac{1}{T^3T_2}
\end{equation}
which holds as long as $T$ is smaller than the coherence time of the signal and $T_2$ the coherence time of the probe.

\section{Conclusion}
\label{sec:conclusion}
 The precision limits typical to classical statistics can be overcome by using quantum metrology protocols. In the absence of environmental noise, so that quantum probes evolve unitarily, the so-called Heisenberg limit can be achieved, i.e., the mean squared error can decrease as fast $1/N^2$ with the number of probes employed --- rather than as $1/N$ characteristic to the standard quantum limit which is dictated by classical statistics. In the presence of noise, quantum probes are subject to environmental fluctuations that will typically hinder the achievable resolution. In the most unfavourable case, uncorrelated noise can constrain the quantum enhancement to a constant factor, and therefore bound the error to the standard asymptotic scaling \cite{Huelga1997, Escher2011, DemkowiczDobrzanski2012}. That is the case of all types of semigroup (time-homogeneous) dynamics that include phase covariant terms, which commute with the system Hamiltonian. Uncorrelated dephasing noise that can be described by a Linbladian master equation is a relevant example of this situation. Remarkably, the standard scaling can be surpassed when the dynamics is no longer ruled by a semigroup and becomes time-inhomogeneous. In this case, the ultimate precision in frequency estimation is determined by the system's short-time behaviour, which when exhibiting the natural Zeno regime leads to an asymptotic resolution beyond the SQL, with a standard deviation scaling as $1/N^{3/2}$. It is important to emphasize that the relevant noise feature dictating the precision is the violation of the semigroup composition law at short timescales, while specific non-Markovianity does not play any specific role as far as the asymptotic scaling is concerned \cite{Smirne2016}.\\
The consideration of specific microscopic models allows for the investigation of the physical mechanisms that lead to a reduction of the attainable precision in a metrology protocol. Using the spin-boson model with weak coupling of arbitrary geometry we can show how imposing the secular approximation leads to a phase-covariant dynamics, while the inclusion of non-secular terms breaks the phase-covariance. In the case of baths with an Ohmic spectral density we can provide an exhaustive characterization of the metrological performance and demonstrate the generality of the Zeno bound beyond phase covariance. Zeno scaling holds unless probes are coupled to the baths in the direction perfectly transversal to the encoding, where a novel scaling proportional to $1/N^{7/4}$ arises \cite{Haase2017}. \\
Many open questions remain to be addressed in the context of open system metrology. We expect that the methods presented in this tutorial can be also useful for the analysis of precision bounds in the small $N$ domain where most practical applications will be developed and where the intricacy of the combination of coherent and incoherent dynamics is expected to be efficiently exploited for achieving super-classical performance.

%\acknowledgements
\section*{Acknowledgments}

This work has received funding
from the Horizon 2020 research and innovation program of the European Union under the QUCHIP project GA no. 641039.
J.K. acknowledges support from the Spanish MINECO (QIBEQI FIS2016-80773-P 
and Severo Ochoa SEV-2015-0522), Fundacio Cellex and Generalitat de Catalunya 
(SGR875 and CERCA Program). R.D.D acknowledges support from National Science Center (Poland) grant No. 2016/22/E/ST2/00559.

\end{document}

%% file: commands.tex
\definecolor{green}{rgb}{0,0.5,0}

\newcommand{\w}{\omega_0}
\newcommand{\R}{\rho^{(N)}_{\w,t}} % state
\newcommand{\chan}{\Lambda^{\otimes N}_{\w,t}} %channel
\newcommand{\Cang}{{\vartheta}} %Coupling angle
\newcommand{\topt}{t_{\mathrm{opt}}} %optimal time
\newcommand{\wc}{\omega_C} %cutoff frequency

\newcommand{\ket}[1]{\left|{#1}\right>}
\newcommand{\bra}[1]{\left<{#1}\right|}
\newcommand{\ketbra}[2]{\left|{#1}\right>\left<{#2}\right|}
\newcommand{\braket}[2]{\left<{#1}\right.\left\vert{#2}\right>}
\newcommand{\sz}{\sigma_z}
\newcommand{\sy}{\sigma_y}
\newcommand{\sx}{\sigma_x}
\newcommand{\id}{\mathds{1}}

\newcommand{\tr}[1]{\mathrm{tr}\left[{#1}\right]}
\newcommand{\ptr}[2]{\mathrm{tr}_{#1}\left[{#2}\right]}
\newcommand{\E}[1]{\left\langle{#1}\right\rangle}
\newcommand{\D}{\mathrm{d}}
\newcommand{\var}[1]{\mathrm{var}\left[{#1}\right]}
\newcommand{\abs}[1]{\left \vert {#1} \right \vert}
\newcommand{\Fcl}{F_{\mathrm{cl}}}